\documentclass{article}
\usepackage{arxiv}
\usepackage{amsthm}
\usepackage{caption}
\usepackage{amsmath, amssymb}
\usepackage{dsfont}
\usepackage{graphics}
\usepackage{graphicx}
\usepackage{bbm}
\usepackage{appendix}
\usepackage{relsize}
\usepackage{xfrac}
\usepackage{cprotect}
\usepackage{longtable}
\usepackage{booktabs}

\usepackage{thmtools}
\usepackage{thm-restate}

\usepackage{hyperref}

\usepackage{cleveref}
\usepackage[graphicx]{realboxes}

\declaretheorem[name=Theorem,numberwithin=section]{thm}
\declaretheorem[name=Corrollary,numberwithin=section]{cor}
\declaretheorem[name=Proposition,numberwithin=section]{prop}
\declaretheorem[name=Lemma,numberwithin=section]{lem}

\newcommand{\bigCI}{\mathrel{\text{\scalebox{1.07}{$\perp\mkern-10mu\perp$}}}}

\newtheorem{assump}{}

\DeclareMathOperator*{\E}{\mathbb{E}}

\usepackage{natbib}
\usepackage{enumitem}

\newtheorem{definition}{Definition}
\usepackage{amsmath}
\usepackage{accents}

\newcommand{\ubar}[1]{\underaccent{\bar}{#1}}
\title{Permutation Weighting}

\author{
    David Arbour\\
    Adobe Research\\
     \texttt{darbour26@gmail.com}\\
     \AND
     Drew Dimmery\\
     Facebook\\
     \texttt{pddimmery@gmail.com}\\
     \AND
     Arjun Sondhi\\
     Flatiron Health
}

\begin{document}

\maketitle

\begin{abstract}
In observational causal inference, in order to emulate a randomized experiment, weights are used to render treatments independent of observed covariates. 
This property is known as balance; in its absence, estimated causal effects may be arbitrarily biased. 
In this work we introduce \textit{permutation weighting}, a method for estimating balancing weights using a standard binary classifier (regardless of cardinality of treatment). 
A large class of probabilistic classifiers may be used in this method; the choice of loss for the classifier implies the particular definition of balance. 
We bound bias and variance in terms of the excess risk of the classifier, show that these disappear asymptotically, and demonstrate that our classification problem directly minimizes imbalance. 
A wide variety of existing balancing weights may be estimated through this regime, allowing for direct comparison between methods based on classifier loss, as well as hyper-parameter tuning using cross-validation. 
Empirical evaluations indicate that permutation weighting provides favorable performance in comparison to existing methods.

\end{abstract}
\section{Introduction}
Observational causal inference methods infer causal effects in the absence of an explicit randomization mechanism.
Given observed treatments, outcomes, and a sufficient set of confounding pretreatment covariates, identification of the causal effect is made possible by rendering treatment independent of the covariates~\citep{rubin2011causal, pearl2009causality}.
Inverse propensity score weighting (IPW) is a common way to accomplish this, where outcomes are weighted by the inverse probability of receiving the observed treatment given covariates~\citep{rosenbaum1983central}.
If these probabilities correctly represent the conditional distribution, then the weighted data will have independence between treatment and covariates.
This property is known as \textit{balance}; for example, in a binary or categorical treatment setting, all treatment groups would have the same weighted distribution of covariates.
Unlike in design-based causal inference, where the relationship between treatment and covariates is known by design, propensity scores often must be estimated from observed data.
Under model misspecification, however, there are no guarantees of balance, and there may remain arbitrary dependencies between treatment and covariates.
Nevertheless, IPW has become widely used in a variety of fields, e.g., epidemiology \citep{cole2008constructing}, economics \citep{hirano2003efficient} and computer science \citep{dudik2011doubly}.

Covariate balancing weights~\citep{hainmueller2012entropy, imai2014covariate, zubizarreta2015stable} seek to remedy the problems of imbalance under misspecification by optimizing a balance condition directly. 
The promise of these techniques is that even when the propensity score model is misspecified, the method will still reduce confounding bias by optimizing for its respective balance condition.
As the balancing weight literature grows, proposed methodologies are differentiated largely by two aspects: (1) the choice of the distance employed as a measure of balance, and (2) the optimization procedure. 
This presents a challenge for practitioners, since the appropriate measure of balance is application-specific and many of the proposed optimization procedures~(e.g. \citet{zubizarreta2015stable, hainmueller2012entropy}), have hyperparameters which must be manually specified and can significantly affect performance. 
In addition, the aforementioned work focuses on the binary treatment regime, but many applied problems are not simple dichotomous treatments.
While \citet{fong2018covariate} provides a linear-balancing method for general treatments, this task still requires hard choices for practitioners about how to specify balance and how to parameterize the conditional distribution of treatment.
Thus, providing a unified framework for comparing balancing weight estimators is critical for effective application. 
\citet{zhao2019covariate} provides one step in this direction by unifying many existing balancing weights for binary treatments by considering proper scoring rules, but does not provide guidance for model selection.

In this paper, we present permutation weighting~(PW), a method for estimating balancing weights for general treatment types by solving a binary classification problem. 
In contrast to prior work, where the target distribution is implicitly defined via the balance objective, PW explicitly represents the balanced dataset by permuting observed treatments, emulating the target randomized control trial (RCT)~\citep{hernan2008does}. 
As a result, the problem of inferring balancing weights reduces to estimating importance sampling weights between the observed and permuted data. 
We estimate these importance sampling weights using classifier-based density ratio estimation~\citep{qin1998inferences, cheng2004semiparametric, bickel2007discriminative}.
This procedure is amenable to general treatment types---binary, multi-valued or continuous---and reduces them all to the same simple binary classification problem which can be solved with off-the-shelf methods.
The choice of classifier and specification of the classification problem implies the balance condition.
Existing methods \citep{imai2014covariate, hazlett2016kernel, fong2018covariate, zhao2019covariate} with balance constraints correspond to particular choices of loss and feature representations for this classifier.
We show that minimizing error in our classification problem directly minimizes the bias and variance of the causal estimator, and imbalance.
This property also implies that cross-validation can be used to tune classifier hyperparameters (section~\ref{subsec:consistency}) and choose between balancing weight specifications using standard software (section~\ref{subsec:scoring_rules}).

To summarize, this paper makes three contributions to the literature on balancing weights:
\begin{enumerate}
    \item We show how to use \textbf{off-the-shelf classifiers} for weight estimation.
    \item We tie causal estimation to classification error (defined through proper scoring rules), providing justification for using \textbf{cross-validation} for hyperparameter tuning and the selection of balance criteria.
    \item The capability to model \textbf{arbitrary treatment types} within the same theoretical and practical framework.
\end{enumerate}

The rest of the paper is structured as follows. 
Section \ref{sec:problem} introduces necessary background and the problem setting of causal inference, balance and weighting methods. 
Section \ref{sec:pw} introduces permutation weighting and in Section \ref{sec:properties} we discuss properties of the method. 
Section~\ref{sec:relatedwork} shows the connections to prior work on balancing weights.
Finally, we evaluate the efficacy of our method for causal inference on binary and continuous treatments in Section \ref{sec:experiments}. 

\section{Problem Statement and Related Work}
\label{sec:problem}
We first fix notation used throughout.
We denote random variables using upper case, constant values and observations drawn from random variables in lower case, and denote a set with boldface.
We will refer to estimates of quantities using hats, e.g., $\hat{w}$ is an estimate of $w$.
Let $\mathcal{D}$ be a dataset consisting of treatments $\textbf{A}$ defined over a domain $\mathcal{A}$, real valued outcomes $Y \in \mathbb{R}$, and a set of covariates $\textbf{X}$ defined over a domain $\mathcal{X}$. 
Note that in our setup, we make no assumption on the cardinality of treatment.
Finally, we denote a \textit{potential outcome} as $Y(\textbf{a})$, which represents an outcome that would have been observed if treatment $\textbf{a}$ had been assigned. 

We assume the following properties of the observed data throughout this work:
\begin{assump}
\label{assump:ignorability}
Ignorability, i.e., $Y(\textbf{a}) \bigCI \textbf{A} \mid \mathbf{X} \quad \forall \textbf{a} \in \mathcal{A}$
\end{assump}
\begin{assump}
\label{assump:positivity}
Positivity over treatment status, i.e., $p(\textbf{A} = \textbf{a} \mid \mathbf{X}=\mathbf{x}) > 0 \quad \forall \mathbf{x} \in \mathcal{X},  \textbf{a} \in \mathcal{A}$
\end{assump}

The causal estimand we focus our attention on is the \textit{dose-response function}, $\E[Y(\textbf{a})]$, i.e., the expected value of the outcome after intervening and assigning treatment to value $\textbf{a}$.
This is a general construct that does not presuppose a specific type, e.g. binary, for treatment. 
Further, the identification of the dose-response function implies identification of many common treatment contrasts of interest.
For example, the average treatment effect under binary treatments $\mathcal{A} = \left\{0, 1\right\}$, is given as $\E[Y(1)] - \E[Y(0)]$.

\subsection{Balance}
\label{subsec:balance}
A common approach to obtain an unbiased estimate of the dose-response function is to render treatments independent from the confounding variables, $\mathbf{X}$~\citep{pearl2009causality, rubin2011causal}.
Within the causal inference literature, this independence is often referred to as the \textit{balance} condition.
We define a general notion of imbalance as some divergence between the observed joint distribution, $p(\mathbf{A}, \mathbf{X})$, and the product distribution, $p(\mathbf{A})p(\mathbf{X})$.

In the binary treatment setting, where balance is most commonly considered, this reduces to performing a two sample test between covariates under treatment and control, i.e. ${D}(\phi(\mathbf{X}_C), \phi(\mathbf{X}_T)) = 0$, where ${D}$ denotes some divergence, $\phi(\cdot)$ is some function, and $\mathbf{X}_C$ and $\mathbf{X}_T$ refer to instances of $\mathbf{X}$ associated with control and treatment, respectively. 
Common balancing weights can be derived with different choices of ${D}$ and $\phi$. 
Section~\ref{sec:relatedwork} walks through these choices for some of these weights.

When treatment is not binary, e.g., continuous or multi-valued, the balance condition must be described explicitly in terms of independence between $\mathbf{A}$ and $\mathbf{X}$, rather than indirectly via the two sample condition. 
While there are a number of definitions, we will focus on divergences which can be described with an $L_p$ norm of the form
\begin{align}
\label{eq:crosscov}
\| \mathbb{E}\left[\phi(\mathbf{X}) \otimes \psi(\mathbf{A})\right] - \mathbb{E}\left[\phi(\mathbf{X})\right] \otimes \mathbb{E}\left[\psi(\mathbf{A})\right]\|_p,
\end{align}
where $\otimes$ is the Kronecker product, $\phi$ and $\psi$ are arbitrary functions, and $p$ is the order of the $L_p$ norm. 
It may help build intuition to note that when $\phi$ and $\psi$ are the identity function and $\mathbf{X}$ is univariate, this value is some norm of the covariance between $\mathbf{X}$ and $\mathbf{A}$.
In a more general setting when the functions are contained in some reproducing kernel Hilbert space, equation $\ref{eq:crosscov}$ with an $L_2$ norm is the Hilbert-Schmidt independence criterion~\citep{GreHerSmoBouSch05}. 

\subsection{Importance Weighting}
\label{subsec:imp_weighting}
A common method for estimating the dose-response function is weighting by the inverse of the conditional probability of receiving treatment given observed covariates, i.e., inverse propensity score weighting (IPW)~\citep{rosenbaum1983central, imai2004causal}.
Weighting by the inverse of this score provides the standard Horvitz-Thompson estimator, which reweights data such that there is no relationship between $\mathbf{A}$ and $\mathbf{X}$, providing identification of causal effects.
This is based on the insight that:
\begin{align*}
   \mathbb{E}[Y(\mathbf{a})] = \mathbb{E}\left[\frac{y_i}{p(\mathbf{a}\mid \mathbf{x}_i)}\right] \quad \forall i : \mathbf{a}_i = \mathbf{a},
\end{align*}
i.e. reweighting individual units provides unbiased estimates of the dose-response surface.
In the binary or categorical treatment case, this allows direct aggregation of effects through a weighted average that is consistent for the dose-response.
In the continuous case, consistency requires approaches such as \citet{kennedy2016doublerobust}, which uses local regression to aggregate units together with similar observed dosage.
To improve efficiency, many practitioners use the \citet{hajek1964} estimator which renormalizes weighted averages based on the sum of the weights rather than the number of units; this improves variance in exchange for a small bias which disappears quickly with increasing sample size.
When the marginal distribution of treatment is far from uniform, both inverse propensity score weighting and the H\'{a}jek estimator can have high variance. 
To remedy this, \citet{robins1997causal} proposed inverse-propensity stabilized weighting~(IPSW) which modifies IPW by placing the marginal density of treatment in the numerator, i.e.,
\begin{align*}
    \mathbb{E}[Y(\mathbf{a})] =\E\left[\frac{y_i p(\mathbf{a})}{p(\mathbf{a}\mid \mathbf{x}_i)}\right] \quad \forall i : \mathbf{a}_i = \mathbf{a}.
\end{align*}

When the conditional distribution has been correctly specified in the propensity score estimation procedure, IPW results in the balance condition~\citep{rosenbaum1983central}, i.e. the weighted distribution of $\mathbf{X}$ is the same for all values of $\mathbf{A}$.
However, when the conditional distribution is \emph{not} well specified, either in terms of the functional form or the assumed sufficient set of pretreatment covariates, inverse propensity score weighting may fail to produce balance on the observed covariates, and the resulting causal estimate may be badly biased ~\citep{harder2010propensity, kang2007demystifying}.

In this work we revisit the the definition of IPSW, providing a previously undocumented identity:
\begin{align}
\label{eq:importancesampler}
\E\left[y_i \frac{p(\mathbf{a}_i)}{p(\mathbf{a}_i\mid \mathbf{x}_i)}\right]
=  \E\left[y_i\frac{p(\mathbf{a}_i)}{\frac{p(\mathbf{a}_i, \mathbf{x}_i)}{p(\mathbf{x}_i)}}\right] 
= \E\left[y_i\frac{p(\mathbf{a}_i)p(\mathbf{x}_i)}{p(\mathbf{a}_i, \mathbf{x}_i)}\right].
\end{align}
This makes plain that the weights given by IPSW define importance sampling weights where the target distribution is the distribution under balance.
To be explicit, the goal of IPSW is to transform expectations over the observed joint distribution of $\mathbf{A}$ and $\mathbf{X}$ to expectations over $\mathbf{A}$ and $\mathbf{X}$ in which they appear as if generated from an RCT (the ``target trial'').
However, the importance sampling weights under IPSW are constructed indirectly by separately estimating the conditional and marginal treatment densities.
The contribution of this work is a method, permutation weighting, which estimates this quantity directly via a probabilistic classification problem which we describe in the next section. 
Direct estimation provides more than just intuitive appeal. 
Unlike IPSW, direct estimation of the importance sampling ratio explicitly seeks to minimize imbalance, which we show in section~\ref{sec:properties}.
We also show that this approach leads to bounds on bias and variance of the dose-response estimates based on classifier error.
The result is that bias is reduced under direct estimation of the density ratio, even in the case of misspecification.

\section{Permutation Weighting}
\label{sec:pw}
We now introduce permutation weighting, which allows for the direct estimation of the importance sampler defined by equation \ref{eq:importancesampler}.
Permutation weighting consists of two steps: 
\begin{enumerate}
    \item The original dataset is stacked with a dataset in which $\mathbf{A}$ has been permuted.
    The permuted dataset is equivalent to fixed-margin randomization of treatment, so represents a distribution where $\mathbf{A}$ and $\mathbf{X}$ are independent. That is, it obeys the balance condition \emph{by design}.
    In what follows, we denote the distribution that the observed data is drawn from as $P$, and the product distribution resulting from permutation as $Q$.
    \item The importance sampling weights, $\hat{w}(\mathbf{a}_i, \mathbf{x}_i)$, are constructed by estimating the density ratio between $P$ and $Q$.
\end{enumerate}
In order to estimate the density ratios~(step 2), we employ classifier-based density ratio estimation~\citep{qin1998inferences, cheng2004semiparametric, bickel2007discriminative}, which transforms the problem of density ratio estimation into binary classification by building a training set from the concatenation of the observed and permuted datasets. 
$\mathbf{A}$ and $\mathbf{X}$ are used as features, and a label, $C \in \{0, 1\}$, is given to denote the membership of the instance to the observed or the permuted dataset, respectively.
A probabilistic classifier learns to recover $p(C=1 \mid \mathbf{A}, \mathbf{X})$.
We denote the true conditional probability as $\eta$ and the estimated conditional probabilities from the classifier as $\hat{\eta}$.
To aid discussion, with some abuse of notation, we will also refer to the classifier which produced the conditional probabilities as $\hat{\eta}$.
After training the classifier, assuming equally sized observed and permuted datasets, the importance weights are recovered by taking the density of the distribution of $\mathbf{A}$ and $\mathbf{X}$ in the permuted dataset ($dQ$) over the density of the observed joint distribution ($dP$)~\citep{bickel2007discriminative}:
\begin{align}
    \nonumber w(\mathbf{a}_i, \mathbf{x}_i) &=
    \frac{\eta(\mathbf{a}_i, \mathbf{x}_i)}
    {1 - {\eta}(\mathbf{a}_i, \mathbf{x}_i)}
    \nonumber = \frac{p(C=1 \mid \mathbf{a}_i, \mathbf{x}_i)}{p(C=0 \mid \mathbf{a}_i, \mathbf{x}_i)} \\
    \nonumber &= \frac{p(C=1, \mathbf{a}_i, \mathbf{x}_i)}{p(C=0, \mathbf{a}_i, \mathbf{x}_i)}\frac{(p(C=1)dQ + p(C=0)dP)}{(p(C=1)dQ + p(C=0)dP)} = \frac{p(C=1, \mathbf{a}_i, \mathbf{x}_i)}{p(C=0, \mathbf{a}_i, \mathbf{x}_i)} \\
    &= \frac{p(\mathbf{a}_i, \mathbf{x}_i | C=1)p(C=1)}{p(\mathbf{a}_i, \mathbf{x}_i |C=0)p(C=0)} =  \frac{dQ}{dP}
    \label{eq:PWIS}
\end{align}
When $dQ$ breaks dependence between $\mathbf{A}$ and $\mathbf{X}$~(e.g. permuting by treatment assignment or specifying the full cross-product), the resulting importance sampler is $\frac{p(\mathbf{a}_i)p(\mathbf{x}_i)}{p(\mathbf{a}_i, \mathbf{x}_i)}$.
The use of a probabilistic classifier for density ratio estimation has a growing literature~\citep{sugiyama2012density, menon2016linking, mohamed2016learning} but it has yet to be employed in the context of observational causal inference.

We define the classifier loss as some function $\lambda: \{0, 1\} \times [ 0, 1 ] \mapsto \mathbb{R}_+$. 
Using $\mathcal{D}$ to denote the distribution of the stacked dataset over which the classifier is trained on (an equal mixture of $P$ and $Q$), the risk for the classifier $\hat{\eta}$ under loss $\lambda$ is then defined as
\begin{align*}
\mathbb{L}(\hat{\eta} ; \mathcal{D}, \lambda) &=\mathbb{E}_{P}\left[\lambda_{1}(\hat{\eta}(a, \mathbf{x}))\right]+\mathbb{E}_{Q}\left[\lambda_{0}(\hat{\eta}(a, \mathbf{x}))\right] 
\end{align*}
The Bayes risk is given as~\citep{reid2011information, menon2016linking}
\begin{align*}
    \mathbb{L}^*(\mathcal{D}, \lambda) = \min_{\hat{\eta}} \mathbb{L} (\hat{\eta}; \mathcal{D}, \lambda)
\end{align*}
The regret is defined as the difference between risk of a classifier, $\hat{\eta}$ and the Bayes-risk,
\begin{align*}
    \operatorname{reg}(\hat{\eta} ; \mathcal{D}, \lambda) =  \mathbb{L}(\hat{\eta} ; \mathcal{D}, \lambda) -  \mathbb{L}^*(\mathcal{D}, \lambda)
\end{align*}

In order to ensure that the probabilities produced by the classifier are well calibrated, we introduce the following assumption:
\begin{assump}
\label{assump:psr}
The classifier, $\hat{\eta}$,
is trained using a twice differentiable strictly proper scoring rule, i.e., $\hat{\eta} \neq \eta \implies \mathbb{L}(\hat{\eta} ; \mathcal{D}, \lambda) > \mathbb{L}(\eta ; \mathcal{D}, \lambda)$~\citep{buja2005loss, gneiting2007strictly}.
\end{assump}

More intuitively, strictly proper scoring rules define functions which, when minimized, provide calibrated forecasts. 
The most common examples of strictly proper scoring rules are logistic, exponential, and quadratic losses~\citep{gneiting2007strictly}.
Strictly proper scoring rules also provide a natural connection to statistical divergences: every proper scoring rule is associated with a divergence between the estimated and true forecasting distribution~\citep{reid2011information, huszarscoring}.
Finally, we also assume consistency of the classifier under the data-generating process. 
\begin{assump}
\label{assump:consist}
The error of the classifier, $\hat{\eta} - \eta$, scales as $O(n^{-\epsilon}), \epsilon \in (0, 1)$.
\end{assump}

If the permuted dataset obeys the balance condition, then the weights will target balance.
In finite samples, this dataset may not have perfect balance, so we perform multiple permutations where the classification procedure is carried out to obtain weights, which are averaged to provide the final estimate of the weight. 
Justification for this procedure is provided in the proof of Corollary~\ref{cor:fixedBalance} in section~\ref{sec:properties}, which relies on the fact that each permutation is a random sample from the ideal balanced distribution.
For low cardinality treatments, the permuted dataset can be constructed as the cross product of the unique values of treatment with $\mathbf{X}$, and no iteration is necessary.

Inferring weights using a classifier confers three important advantages:
\begin{enumerate}[leftmargin=*]
\item Regardless of the type of the treatment~(binary, continuous, multinomial, etc.), the problem reduces to the same binary classification task. 
In contrast to many existing methods, this means that it is not necessary to explicitly assume a parametric form for the treatment conditional on covariates (for instance, generalized propensity scores often assume that dosage is conditionally normal).
As such, the use of binary classification to directly estimate weights requires weaker assumptions in environments with complicated treatments.
\item As we discuss in section \ref{sec:properties}, minimizing the error of the binary classifier directly results in minimizing both imbalance~(proposition \ref{prop:balance}) and the error of the causal estimate itself~(propositions \ref{prop:bregmanBoundedBias} and \ref{prop:bregmanBoundedVariance}).
As a result, both the hyperparameters and the measure of balance itself~(via the choice of feature representation and loss) can be optimized directly by considering the cross-validated error of the binary classifier. 
In addition to theory, we demonstrate the empirical efficacy of hyperparameter tuning and model selection for estimating causal effects in section \ref{subsec:crossvalexps}.
\item There is a deep connection between binary classification and two sample testing~\citep{friedman2004multivariate, reid2011information}. 
Through this lens, the choice of feature representation and classification loss is equivalent to choosing a balance condition. 
We provide an examination of the relationship between permutation weighting and a number of existing weighting estimators~\citep{hainmueller2012entropy,  imai2014covariate,  hazlett2016kernel, zhao2019covariate}.
We are explicit about this in Section~\ref{sec:relatedwork}, showing that many of these existing approaches are expressable under the framework of permutation weighting.
\end{enumerate}

\section{Properties}
\label{sec:properties}

We now examine the finite sample and asymptotic behavior of permutation weighting.
To do so, we will first consider a slightly more general setting than the procedure outlined in the previous section. 
Specifically, propositions \ref{prop:bregmanBoundedBias}, \ref{prop:bregmanBoundedVariance}, \ref{prop:consistency} and \ref{prop:balance} examine the behavior of importance sampling from the observed distribution $P$ to an arbitrary distribution $Q$ (under positivity, assumption \ref{assump:positivity}), using a classifier trained with a strictly proper scoring rule (assumption \ref{assump:psr}). 
These may be of independent interest as they admit reasoning over a broad class of estimands~\citep[e.g.][]{bickel2007discriminative, sugiyama2012density, menon2016linking}, including common causal estimands like the average treatment effect on the treated.
Indeed, any distribution of $\mathbf{A}$ and $\mathbf{X}$ which conforms to the overlap assumption can be used as a target distribution under this framework.
Before presenting our results, we first introduce Bregman divergences, a class of statistical distances.
\begin{definition}[Bregman divergence~\citep{bregman1967relaxation}]
Define the Bregman generator, $g : S \to \mathbb{R}$, to be a convex, differentiable function.
The difference between the value of $g$ at point $s$ and the value of the first-order Taylor expansion of $g$ around point $s_0$ evaluated at point $s$ is given by
$B_g(s, s_0) \equiv g(s) - g(s_0) - \langle s - s_0, \nabla g(s_0) \rangle$.
\end{definition}
Minimizing many commonly used classification losses correspond to minimizing a Bregman divergence, e.g., accuracy (0-1) loss corresponds to total variation distance, squared loss corresponds to triangular discrimination distance, log loss corresponds to the Jensen-Shannon divergence, and exponential loss corresponds to the Hellinger distance~\citep{reid2011information}. 
The latter three losses are proper scoring rules and conform to our assumption~\ref{assump:psr}.
All strictly proper scoring rules have a corresponding Bregman divergence~\citep{dawid2007geometry}.
We use this correspondence in the proofs of our theoretical results, which are generally deferred to the supplement. 

\subsection{Estimation}
\label{subsec:consistency}
Throughout this section we will denote the target weights for the permutation weighting importance sampler as $w$ and the estimated weights $\hat{w}$.
We now begin by deriving bounds on the bias for weighting estimators.
\begin{restatable}[Bias of PW]{prop}{pwbias}
\label{prop:bregmanBoundedBias}
Let $\mathbb{E}_P$ and $\mathbb{E}_Q$ denote the expectation under the distributions $P$ and $Q$, respectively. 
The bias of the dose response function $\mathbb{E}_P[y\hat{w}]$ with respect to $\mathbb{E}_Q[y]$ is bounded by
    $\left|\mathbb{E}_{Q}[y] -  \mathbb{E}_{P}\left[y \hat{w}\right]
    \right| \leq \mathbb{E}_{P}\left[\frac{2|y|}{\sqrt{{g''(1)}}} \sqrt{\operatorname{reg}(\hat{\eta} ; \mathcal{D}, \lambda)}\right]$, 
where $g(\cdot)$ is a Bregman generator. 
\end{restatable}

Minimizing this bound corresponds to minimizing the regret of the classsifier.
We next bound the variance of the permutation weighting dose-response estimator. 
\begin{restatable}[Variance]{prop}{pwvariance}
\label{prop:bregmanBoundedVariance}
Let $\mathbb{V}_{Q}[y]$ denote the variance of $Y$ under the distribution $q$. 
$\mathbb{V}_{Q}[y]$ is bounded by
$$\mathbb{V}_{Q}[y] \leq \frac{1}{n}\mathbb{E}_{Q}[y^2] + 
\frac{4}{n{\sqrt{{g''(1)}}}} {\sqrt{\operatorname{reg}(\hat{\eta} ; \mathcal{D}, \lambda)}} \mathbb{E}_P \left[{y^2}w  + \frac{{y^2}}{\sqrt{g''(1)}} \sqrt{\operatorname{reg}(\hat{\eta} ; \mathcal{D}, \lambda)}\right]$$.
\end{restatable}

The bounds given by propositions \ref{prop:bregmanBoundedBias} and \ref{prop:bregmanBoundedVariance} demonstrate that the quality of importance sampling weights is governed by the regret of the classifier used.
For KL divergence, the bound given by proposition~\ref{prop:bregmanBoundedBias} is essentially Pinsker's inequality~\citep{reid2010composite}.

Finally, consistency of the importance sampler used by permutation weighting is given by the following proposition, which follows as a consequence of propositions \ref{prop:bregmanBoundedBias} and \ref{prop:bregmanBoundedVariance}:
\begin{restatable}[Consistency]{prop}{pwconsistency}
\label{prop:consistency}
Under Assumptions \ref{assump:ignorability}-\ref{assump:consist}, and bounded outcomes $y$, the permutation weighting dose-response estimator is consistent, i.e., as $n \longrightarrow \infty$, $\mathbb{E}_P [y\hat{w}]  \longrightarrow \mathbb{E}_{Q}[y]$.
\end{restatable}

\subsection{Balance}
\label{subsec:balance_properties}

We next show how the importance sampler provided by permutation weighting provides balance.
We preface this with a general definition of balance:

\begin{definition}[Functional discrepancy]
The $L_p$ functional discrepancy for functions $\phi$ and $\psi$, under a weighting estimator $\hat{w}$ is
\begin{align*}
 &\left\|\mathbb{E}_P \left[\phi(\mathbf{a}_i)\otimes\psi(\mathbf{x}_i) \hat{w}(\mathbf{a}_i, \mathbf{x}_i)\right] - \mathbb{E}_Q\left[\phi(\mathbf{a}_i)\otimes\psi(\mathbf{x}_i)\right] \right\|_p.
 \end{align*}
 When the target distribution can be factored as $p(\mathbf{A})p(\mathbf{X})$, this is a measure of imbalance:
 \begin{align*}
 \left\|\mathbb{E}_P\left[\phi(\mathbf{a}_i)\otimes\psi(\mathbf{x}_i) \hat{w}(\mathbf{a}_i, \mathbf{x}_i)\right] - \mathbb{E}_{P}\left[\phi(\mathbf{a}_i)\right]\otimes\mathbb{E}_{P}\left[\psi(\mathbf{x}_i)\right] \right\|_p.
\end{align*}
\end{definition}

This quantity is the extent to which the reweighted expectation differs from the expectation under the product distribution.
A functional discrepancy of zero \emph{for all} $\phi$ and $\psi$ implies independence between $\mathbf{A}$ and $\mathbf{X}$.
When both $\phi$ and $\psi$ are the identity function, then a discrepancy of zero is synonymous with linear balance.
With this definition in hand, we can provide an explicit expression for the functional imbalance attained by permutation weighting:

\begin{restatable}[Minimizing Imbalance]{prop}{pwbalance}
\label{prop:balance}
The $L_p$ functional discrepancy between the observed data drawn from $p(\mathbf{a},\mathbf{x})$ and the proposed distribution $q(\mathbf{a}, \mathbf{x})$ under permutation weighting is
$$\left\|\mathbb{E}_{p(\mathbf{a}, \mathbf{x})}\left[\phi(\mathbf{a}_i)\otimes\psi(\mathbf{x}_i)(\hat{w}(\mathbf{a}_i, \mathbf{x}_i) - w(\mathbf{a}_i, \mathbf{x}_i))\right]\right\|_p
 \leq \frac{2}{\sqrt{{g''(1)}}} \sqrt{\operatorname{reg}(\hat{\eta} ; \mathcal{D}, \lambda)}\left\|\mathbb{E}_{p(\mathbf{a}, \mathbf{x})}\left[\phi(\mathbf{a}_i)\otimes\psi(\mathbf{x}_i)\right]\right\|_p$$
 where $p \geq 0$.
\end{restatable}
Proposition \ref{prop:balance} demonstrates the balancing behavior of the importance sampler employed by permutation weighting. 
The importance sampler defined by permutation weighting has a \textit{linear} dependence on the error of the density ratio estimate which is minimized as classifier regret is minimized.

The above demonstrates properties for estimating importance weights to any pre-specified joint distribution, $Q$, of $\mathbf{A}$ and $\mathbf{X}$.
We now focus our attention on the distribution $p(\textbf{A})p(\mathbf{X})$ -- that of marginal-preserving independence between treatment and covariates.
With a low cardinality treatment (such as binary), it's possible to directly construct a balanced dataset to satisfy $p(\textbf{A})p(\mathbf{X})$.
This is done by taking the cross-product of the unique values of $\textbf{A}$ and of $\textbf{X}$.
Rows can then be weighted to match the marginals from the original data. 
Each row in the pseudo-dataset, $i$, would receive a weight of $\frac{1}{n^2} n(\textbf{a}_i) n(\textbf{x}_i)$, where $n(\textbf{a}_i)$ denotes the number of rows in the original dataset with treatment level $\textbf{a}_i$, and likewise for $n(\textbf{x}_i)$.
However, if $\textbf{A}$ has larger cardinality, it may be computationally difficult or (in the case of continuous treatments) impossible to construct such a dataset.
Instead, the easiest way to target this distribution is through a simple permutation, in which the treatment vector is reshuffled.
By design, this permuted dataset will have no systematic relationship between $\mathbf{A}$ and $\mathbf{X}$ except for that which occurs by chance. 
This is true for the same reason that fixed-margins randomization in RCTs attains balance at expectation.
This easy permutation construction allows the application of all the properties of density ratio estimation discussed above.
Asymptotically in $n$, a single permutation will (like data observed from an RCT) converge to the appropriate balanced target distribution.
In finite samples, there may remain minor imbalances from a single permutation.
For this reason, we propose averaging across multiple permutations to attain an effective balancing weight.

\begin{cor}
\label{cor:consistBalance}
With a single permutation as described above, as $n \to \infty$, and assuming the classifier is consistent, the functional imbalance is minimized.
\end{cor}

\begin{proof}
The weight vector that minimizes
\begin{align*}
\left\|\mathbb{E}_P \left[\phi(\mathbf{a})\otimes\psi(\mathbf{x}) w \right] - \mathbb{E}_{P}\left[\phi(\mathbf{a})\right]\otimes\mathbb{E}_{P}\left[\psi(\mathbf{x})\right] \right\|_p,
\end{align*}
is the density ratio $w$. 
By definition, a single permutation $(\mathbf{a}_{\boldsymbol{\pi}}, \mathbf{x})$ is drawn from the product distribution $Q(\mathbf{a}, \mathbf{x}) = P(\mathbf{a})P(\mathbf{x})$.
Therefore, the trained classifier $\hat{\eta}(\mathbf{a}, \mathbf{x})$ is estimating the probability of $(\mathbf{a}, \mathbf{x}) \sim Q$ instead of $(\mathbf{a}, \mathbf{x}) \sim P$. 
As $n \to \infty$ we have that the error of the classifier probability $\hat{\eta}$ tends to zero by assumption. 
Then, by proposition 6, we have that $\hat{w}(\mathbf{a}_{\boldsymbol{\pi}}, \mathbf{x}) \to w$.
\end{proof}

\begin{cor}
\label{cor:fixedBalance}
For fixed $n$, as the number of permutations increases, the functional imbalance is minimized.
\end{cor}

\begin{proof}
For each $j = 1, \dots, B$, where $B$ is the number of permutations, consider the empirical imbalance minimization problem:
\begin{align*}
    &\min_{\hat{w}_j} \left\|\sum_{i=1}^n \phi(\mathbf{a}_i)\otimes\psi(\mathbf{x}_i) \hat{w}_j(\mathbf{a}_i, \mathbf{x}_i) - \phi(\mathbf{a}_{\boldsymbol{\pi_j}(i)}) \otimes \psi(\mathbf{x}_i) \right\|_p.
\end{align*}
From proposition \ref{prop:balance}, we know that each imbalance is minimized by minimizing the error of the classifier trained in permutation weighting.
Since each permutation $\pi_j$ is drawn from the $n-$sample product distribution $Q_n(\mathbf{a}, \mathbf{x}) = P_n(\mathbf{a})P_n(\mathbf{x})$, as $B \to \infty$, the empirical distribution of $\hat{w}_j$ tends to the distribution of weights computed under this distribution. 
Therefore, the empirical average of weight vectors tends to the weight vector that minimizes  
\begin{align*}
    &\min_{\hat{w}} \int_{(\mathbf{a'}, \mathbf{x'}) \sim Q_n} \left\|\sum_{i=1}^n \phi(\mathbf{a}_i)\otimes\psi(\mathbf{x}_i) \hat{w}(\mathbf{a}_i, \mathbf{x}_i) - \phi(\mathbf{a'}) \otimes \psi(\mathbf{x'}) \right\|_p.
\end{align*}
\end{proof}

\subsection{Choosing among scoring rules}
\label{subsec:scoring_rules}

In contrast to existing work which requires a priori specification of the balance condition, the choice of the balance condition can be performed by considering the out of sample performance of the classifier with respect to the receiver operator characteristic~(ROC) curve, a common measure of classifier performance. 
From the bounds given in propositions~\ref{prop:bregmanBoundedBias} and \ref{prop:bregmanBoundedVariance}, we can select the condition which minimizes the error of the causal estimate.
By noting the connection between the choice of classifier loss and two sample discrepancies provided by \citet{reid2011information}, this procedure also corresponds to choosing a balance condition.
This brings us to the following result from \citet{menon2016bipartite}:

\begin{prop}\label{prop:rocdominance}[ROC dominance: Proposition 13 of~\citep{menon2016bipartite}]
Stochastic dominance of the ROC curve for one classifier over another implies dominance with respect to any strictly proper scoring rule.
\end{prop}

That is, when two ROC curves do not cross, the one with higher true positive rates across all false negative rates will also be superior according to \emph{all} proper scoring rules.
This property shows how the AUC is an effective diagnostic to choose between classifiers.
When the ROC curves for two classifiers cross one another, it may be the case that different choices of loss would suggest different ``optimal'' classifiers.
Within the context of this paper, these results imply that the choice of a balance criterion can be made by examining the ROC curves produced by different modeling choices.
These modeling choices correspond to the estimation of different balancing weights. 

\section{Connections to Prior Work}
\label{sec:relatedwork}
An additional benefit of viewing balancing weight estimation as classification between samples from the observed joint and independent distributions is that it provides the ability to compare a number of existing weighting estimators through the corresponding loss and feature representation choices for the entailed classifier. 

For example, in the setting of binary treatments: 
\begin{itemize}
\item The just-identified condition for the covariate balancing propensity score~\citep{imai2014covariate} is recovered within the permutation weighting setup by considering a polynomial feature expansion and log-loss (derived in Appendix~\ref{app:cbps_connection}).
\item Entropy balancing is reproduced by placing a entropy penalty on a logistic regression~\citep{hainmueller2012entropy, josey2020framework}.
\item Kernel balancing weights~\citep{hazlett2016kernel} can be recovered by considering a Rocchio approximation to a support vector machine using a reproducing kernel Hilbert space as the feature representation. 
\item As first pointed to by \citet{zhao2019covariate}, stable balancing weights are approximated by considering a LASSO style penalty with a polynomial feature expansion. 
\end{itemize}
The quality of the approximation is determined by the extent to which the permuted sample obeys independence between treatment and outcome. 
Full proofs are provided in the appendix, but intuitively the reductions work as follows: the independence criterion imposed by permutation weighting reduces to minimizing a two sample test, or equivalently, the discrepancy between treatment and control covariates~\citep{reid2011information}.
Thus, the specific balance condition being targeted reduces to choices of feature representation and classifier loss.
Note that this reduction is similar to those found in \citet{zhao2019covariate}, where the authors also consider a framework using proper scoring rules. 
We note that there are two central difference in our work.
First, under the proposed framework both the hyperparameters and model selection can be chosen using metrics obtained via cross-validation~(see propositions \ref{prop:bregmanBoundedBias} and \ref{prop:rocdominance}).
In addition to this being appealing theoretically, we see empirically in section \ref{subsec:crossvalexps} that it allows practitioners the ability to plainly reason over what are often complex and difficult decisions. 
Second, the exact same framework applies to arbitrary treatment types, e.g. continuous and multi-valued. 
For example, targeting linear balance in the continuous treatment setting is achieved using the same logistic regression problem that is entailed by just-identified covariate balancing propensity score. 

In addition to being able to reproduce existing balancing weights using permutation weighting, the converse---mapping other weights into our classification setup---is also possible. 
To do this, simply note that a weight, $w$, implies the conditional class probability $p(C=1\mid \mathbf{X}, \mathbf{A}) = \frac{w}{w+1}$~\citep{menon2016linking}.
This allows practitioners to use model diagnostics under the permutation regime to compare weights estimated under different frameworks (e.g. comparing a random forest permutation weight to stable balancing weights or vanilla IPSW).
By examining the ROC curves associated with these weights, they may be compared through a unified framework, wherein ROC dominance implies dominance with respect to any proper scoring rule (proposition~\ref{prop:rocdominance}).

\section{Experiments}
\label{sec:experiments}
For the following simulation studies, we will examine only performance of simple weighting estimators for scalar-valued treatments, 
$E[Y(a)] \approx \sum_{i=1}^n y\hat{w}(a_i, \mathbf{x}_i) K(a_i, a)$. 
For binary treatments, $K(\cdot, \cdot)$ is an indicator for treatment status, while for continuous treatments, it is a kernel weighting term as analyzed in the context of doubly-robust estimators by~\citet{kennedy2016doublerobust}.
This simple estimator is used in our evaluation to provide the most direct test of the efficacy of the various estimators of the weights.
Appendix~\ref{app:more_sims} provides results for the doubly-robust estimators of \citet{kennedy2016doublerobust} as well as the estimation of weighted outcome regressions.
These more complex evaluations do not differ in their substantive conclusions (i.e. rank-order and relative performance).
Error is measured via integrated root mean squared error (IRMSE) as in \citet{kennedy2016doublerobust}, with $s$ indexing $S$ simulations and $\theta_s(a)$ being the unconditional expectation of a given potential outcome in a single simulation, $\E_s[Y(a)]$, i.e.,
$$
\widehat{\text{IRMSE}} = \int_{\mathcal{A}^*}\left[ \frac{1}{S} \sum_{s=1}^S \{\hat{\theta}_s(a) - \theta_s(a)\}^2\right]^{\frac{1}{2}} p(a) da
$$
That is, we take an average of RMSE weighted over the marginal probability of treatment. 
Following \citet{kennedy2016doublerobust}, we perform this evaluation over $\mathcal{A}^*$, the central 90\% of the distribution of $A$ (in the case of binary treatments, we evaluate over the entire support of $A$).
We also evaluate the Integrated Mean Absolute Bias, which replaces the inner average with $\left|\frac{1}{S} \sum_{s=1}^S \{\hat{\theta}_s(a) - \theta_s(a)\}\right|$.
When permutation weighting is used, we perform 100 independent iterations of the permutation procedure to generate weights.
Our evaluations center around two main classifiers: logistic regression and gradient boosted decision trees.
The former focuses on minimizing a log-loss and therefore the balance condition corresponds to minimization of the Jensen-Shannon divergence.
The boosting model corresponds to an exponential loss \citep{lebanon2002boosting} which implies the minimization of the Hellinger divergence.
Achieving equivalence to linear balancing methods using the permutation weighting framework entails the addition of an interaction term between ${A}$ and $\textbf{X}$ due to the different setup of the classification problem; otherwise, the linear classifier would only be able to account for differing marginal distributions of ${A}$ and $\textbf{X}$ (asymptotically, there are no such differences).
We include this interaction in all of the models we evaluate.

\subsection{Binary treatment simulation}
\label{subsec:kangschafer}

Our first simulation study follows the design of \citet{kang2007demystifying}. 
In this simulation, four independent, standard normal covariates are drawn.
A linear combination of them is taken to form the outcome and treatment process (the latter passed through an inverse-logistic function).
Two versions of this standard simulation are analyzed.
One in which the four covariates are observed directly, and one in which only four non-linear and interactive functions of them are observed.
The data is simulated according to the following:
\begin{align*}
X_k &= \mathcal{N}(0,1) \quad \forall k \in \{1,2,3,4\}\\
p(A=1\mid X) &= \textsf{logit}^{-1}\{X_1 - 0.5 X_2 + 0.25 X_3 + 0.1 X_4\}\\
\E[Y\mid A, X] &= 210 + A + 27.4 X_1 + 13.7 X_2 + 13.7 X_3 + 13.7 X_4\\
Y\mid A, X &= \mathcal{N}(\E[Y\mid A, X], 1)
\end{align*}
Under the non-linear "misspecification", the covariates are not observed directly, but instead only the following transformations:

\begin{align*}
X_1&=\exp\left\{\frac{X_1}{2}\right\} & X_2&=\frac{X_2}{1+\exp\{X_1\}}+10 &
X_3&=\left(\frac{X_1 X_3 }{25}+.6\right)^3 & X_4&=(X_2+X_4+20)^2
\end{align*}

\begin{figure}
    \centering
    \includegraphics[width=.8\textwidth]{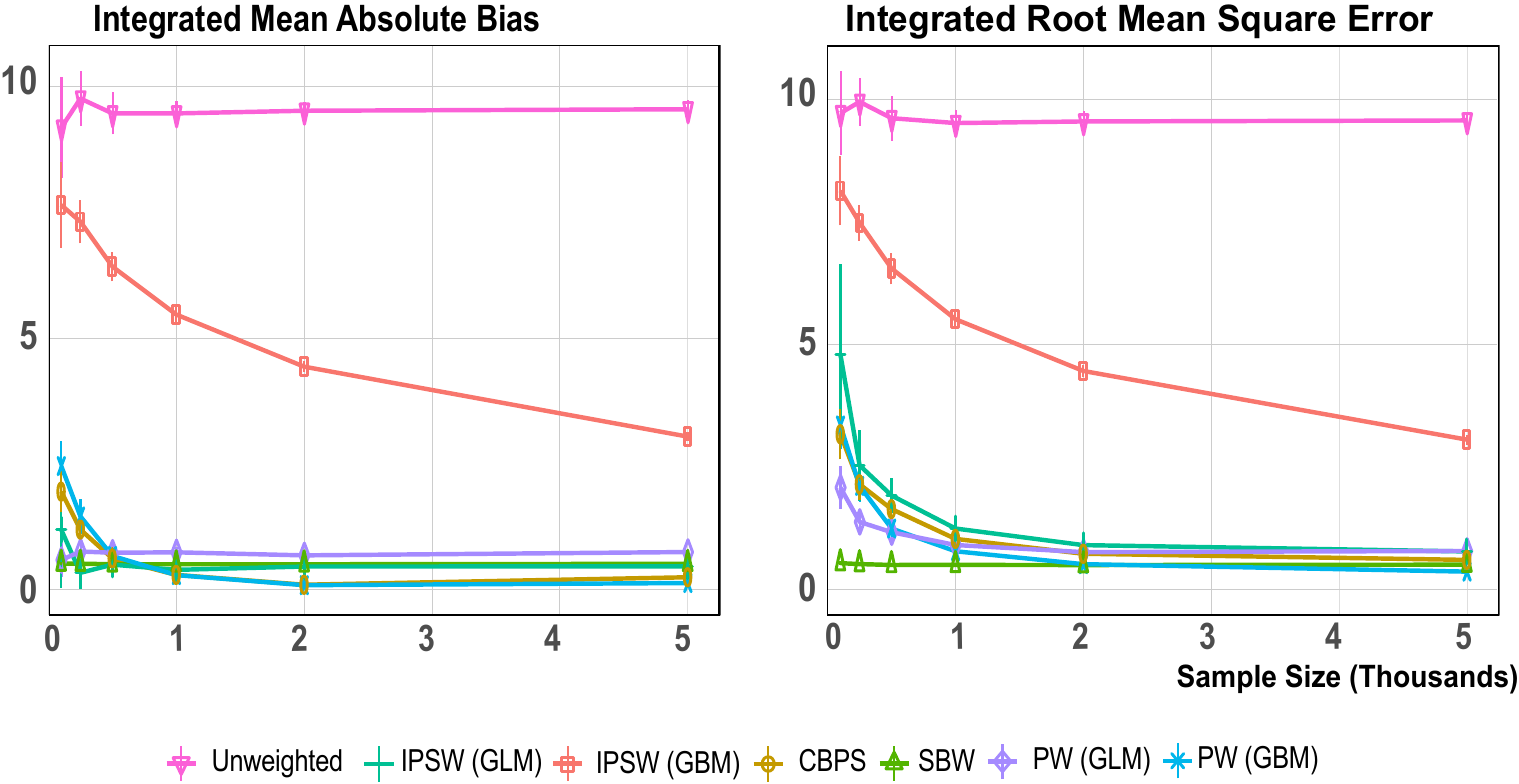}
    \caption{
    Kang-Schafer simulation under correct specification of confounding variables. 
    \texttt{Unweighted} uses no weighting.
    \texttt{IPSW (GLM)} is a logistic-regression-based propensity score model.
    \texttt{IPSW (GBM)} is a propensity score model trained with a gradient boosted decision tree. 
    \texttt{CBPS} is covariate balancing propensity scores \citep{imai2014covariate}.
    \texttt{SBW} is stable balancing weights \citep{zubizarreta2015stable}.
    \texttt{PW (GLM)} is a permutation weighting model using a logistic regression.
    \texttt{PW (GBM)} is a permutation weighting model using a gradient boosted decision tree.
    }
    \label{fig:KS-correct}
\end{figure}

Results from simulations based on the correctly specified data generating process are shown in figure \ref{fig:KS-correct}.
The correct treatment and outcome models are linear in this simulation.
As such, the propensity score model is specified correctly and therefore performs well.
Stable balancing weights also perform very strongly in this case (particularly in low sample sizes), as they both explicitly reduce the variance of the weights and correctly specify the form of the confounding relationship.
These results show that, given sufficient sample size, PW with a logistic regression classifier (or boosting) replicates and eventually outperforms stable balancing weights, even when the true relationships are linear. 
Note, however, that bias under the boosted model is driven to the minimum very quickly, even though the more complicated model increases the variance (hurting the overall IRMSE).
PW with logistic regression typically outperforms CBPS in this simulation, particularly at low sample sizes, despite the fact that both seek to balance the correct specification of confounding.
This difference in performance comes from the minor data-dependent regularization in the score condition estimated by permutation weighting (details in appendix~\ref{app:cbps_connection}).
Machine learning the propensity score performs consistently poorly relative to all other methods examined here, as it imposes no structure on the data.
This demonstrates the importance of seeking balance rather than seeking correct specification.
The latter is very hard; while the former is achievable.

\begin{figure}
    \centering
    \includegraphics[width=.8\textwidth]{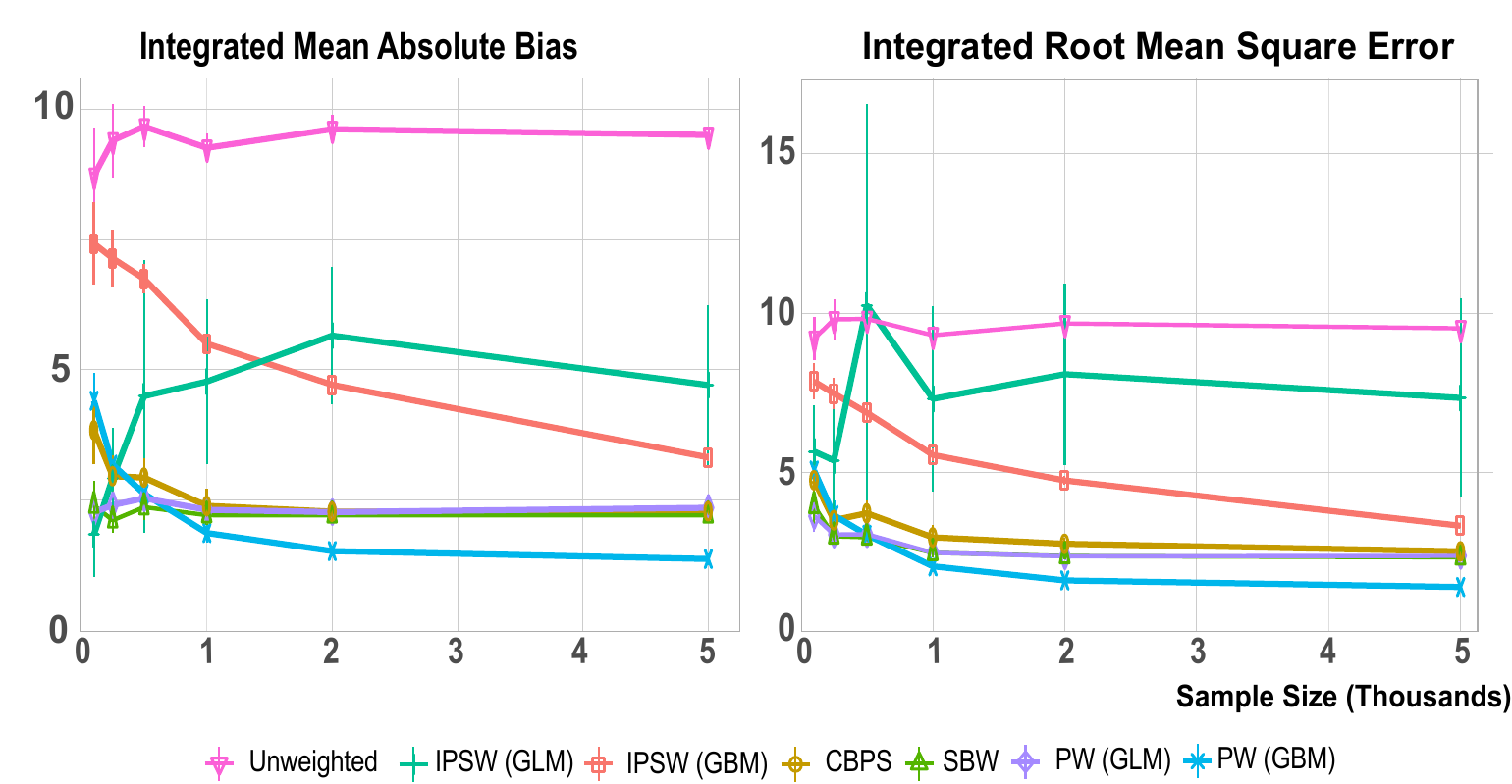}
    \caption{
    On the \citet{kang2007demystifying} simulation under misspecification of confounding variables, PW attains state-of-the-art performance in both bias and RMSE by $n=500$ and reduces RMSE by $30\%$ at $n=2000$.
    \texttt{Unweighted} uses no weighting.
    \texttt{IPSW (GLM)} is a logistic-regression-based propensity score model.
    \texttt{IPSW (GBM)} is a propensity score model trained with a gradient boosted decision tree. 
    \texttt{CBPS} is covariate balancing propensity scores \citep{imai2014covariate}.
    \texttt{SBW} is stable balancing weights \citep{zubizarreta2015stable}.
    \texttt{PW (GLM)} is a permutation weighting model using a logistic regression.
    \texttt{PW (GBM)} is a permutation weighting model using a gradient boosted decision tree.
    }
    \label{fig:KS-incorrect}
\end{figure}

Figure \ref{fig:KS-incorrect} shows results for the more realistic case in which the researcher does not know the correct specifications of the confounding relationships of the covariate set with treatment. 
In these results, PW with boosting drastically improves on the existing weighting estimators, reducing by around $30\%$ the IRMSE relative to balancing propensity scores at $n=2000$.
At smaller sample sizes, the improvements are less substantial, but even by $n=500$, PW with boosting provides superior performance.
This is unsurprising, given that boosting is able to learn a more expressive balancing model (and, therefore, reduce bias) more effectively than other balancing methods.
A standard propensity score estimated by gradient boosted decision trees does not solve the issues faced by propensity scores, leading to large biases in estimation and subsequently large IRMSE across all sample sizes.
Detailed results in tabular format are available in Appendix~\ref{app:more_sims}.
A similar simulation study on a continuous treatment is detailed in Appendix~\ref{app:kangschafer-continuous}.

\subsection{Lalonde evaluation with continuous treatment}
To explore the behavior of permutation weighting under continuous treatment regimes with irregularly distributed data, we turn to the data of \citet{lalonde1986evaluating}, and in particular, the Panel Study of Income Dynamics observational sample of 2915 units (discarding all treated units from the sample and retaining only the experimental control units).
Our evaluation is based around the differences between the experimental control group and the observational control group, which are known to differ greatly based on observed covariates~\citep{smith2005does}.
The covariates in this data are highly non-ellipsoidal, consisting of point-masses and otherwise irregular distributions.
Following the simulation study in \citet{diamond2013genetic}, we simulate a nonlinear process determining assignment of units to dosage level and, then, to outcome based on observed covariates.
The treatment process is made to behave similarly to real-world data by estimating a random forest to predict presence in the experimental / observational sample as a function of observed covariates.
Predicted values from this model are denoted as $\hat{g}(X)$.
Dosage is then a quartic function of that predicted score as well as the nonlinear function determining treatment assignment in \citet{diamond2013genetic}.
The shape of the true dose-response function is similarly a quartic function of dose.

\begin{align*}
	\tilde{P} &= \text{Natural spline basis expansion of }\hat{g}(X) \\
	\boldsymbol \beta &\approx [-0.502\;  0.132\; -0.079\;  0.887]\\
    \mathbb{E}[A \mid X] &= 1 + \tilde{P} \beta + 0.01 * \text{age}^2 
    - .3 \text{education}^2 - 0.01 \log(\text{income74} + .01)^2 + 0.01 \log(\text{income75} + 0.01)^2 \\
    \mathbb{V}[A\mid X] &= 10^2\\
    \tilde{A} &= \text{Natural spline basis expansion of }A\\
    \boldsymbol \gamma &\approx [0.117\;  0.319\; -0.582\;  0.715]\\
    \mathbb{E}[Y\mid A, X] &= \text{income76} + \tilde{A} \gamma
    + .1 \exp\left\{.7 \log(\text{income74} + 0.01) + .7 \log(\text{income75} + 0.01)\right\}\\
    \mathbb{V}[Y\mid A, X] &= 10^2
\end{align*}

\begin{figure}
    \centering
    \includegraphics[width=.78\textwidth]{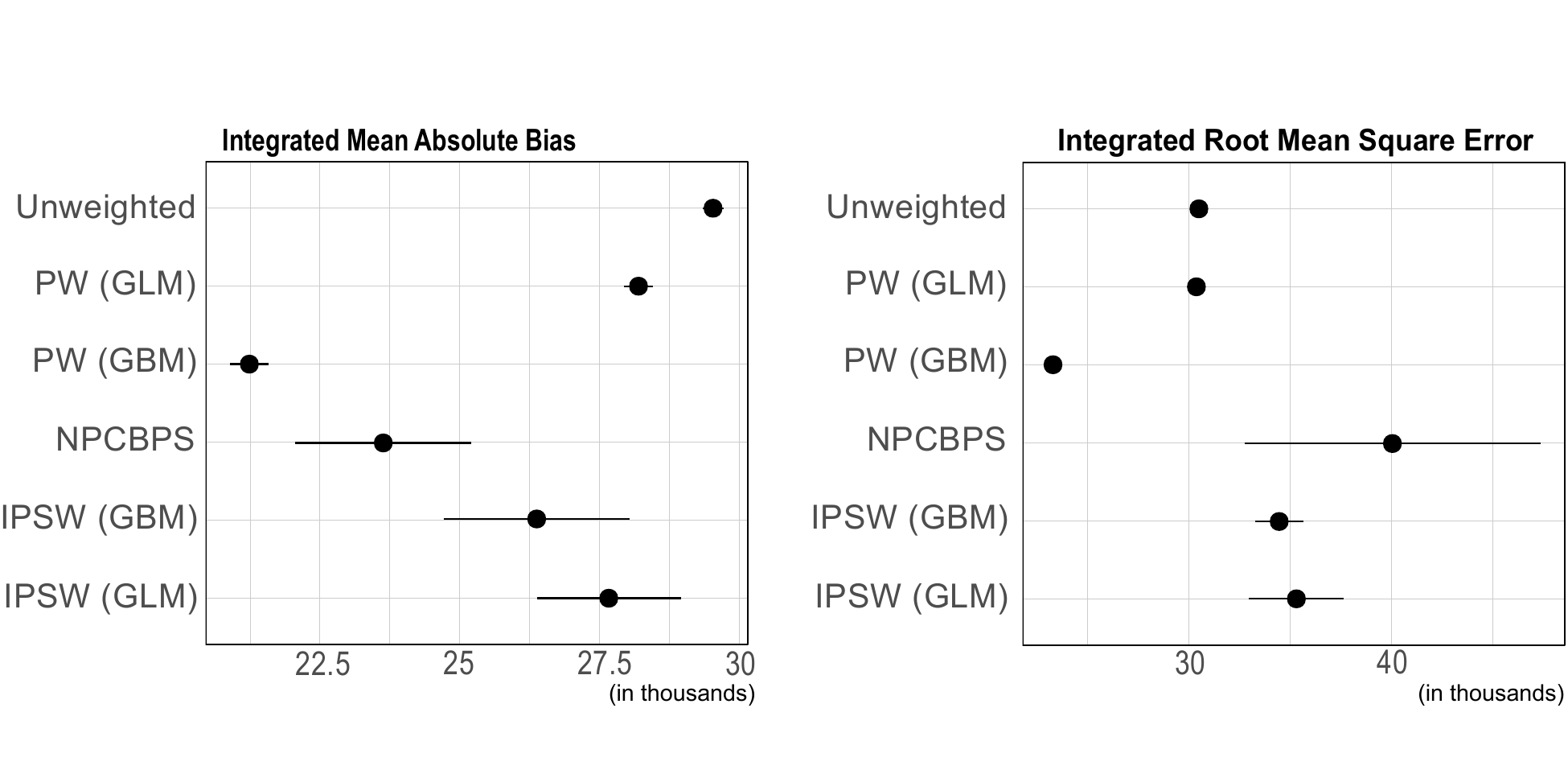}
    \caption{
    On the continuous-treatment simulation based on the LaLonde data, PW greatly reduces both bias and variance relative to existing methods.
    \texttt{Unweighted} uses no weighting.
    \texttt{PW (GLM)} is a permutation weighting model using a logistic regression.
    \texttt{PW (GBM)} is a permutation weighting model using a gradient boosted decision tree.
    \texttt{NPCBPS} is non-parametric covariate balancing propensity scores \citep{fong2018covariate}.
    \texttt{IPSW (GLM)} is a normal-linear regression based propensity score model.
    \texttt{IPSW (GBM)} is a gradient boosted regression generalized propensity score model with homoskedastic normal conditional densities.
    }
    \label{fig:lalonde_sim}
\end{figure}
Figure~\ref{fig:lalonde_sim} shows the IRMSE of a variety of weighting estimators on this simulated benchmark.
Only weights generated by permutation weighting out-perform the raw, unweighted data in terms of IRMSE.
All models reduce bias relative to the raw, unweighted dose-response, but induce unacceptably large variance as they do so.
When a logistic regression is used as the classifier, PW performs better than no weighting by just half a percent in terms of IRMSE (as it does not greatly reduce bias).
When a boosted model is used, however, this gap grows substantially, with PW outperforming the raw estimates by around 25\% -- the only substantial improvement in accuracy among these estimators.
As the earlier simulations have shown, using a boosting model to estimate a standard propensity score does not perform well, increasing IRMSE relative to the unweighted estimate.
It's also worth noting the much reduced variability around the estimates of IRMSE from the permutation weighting models relative to other methods which often have very unstable performance characteristics.
Rank ordering among methods remains largely unchanged when an outcome model is incorporated  (see appendix~\ref{app:lalonde_extra}).

\subsection{Cross-validation}
\label{subsec:crossvalexps}

\begin{figure}
    \centering
    \includegraphics[width=\textwidth]{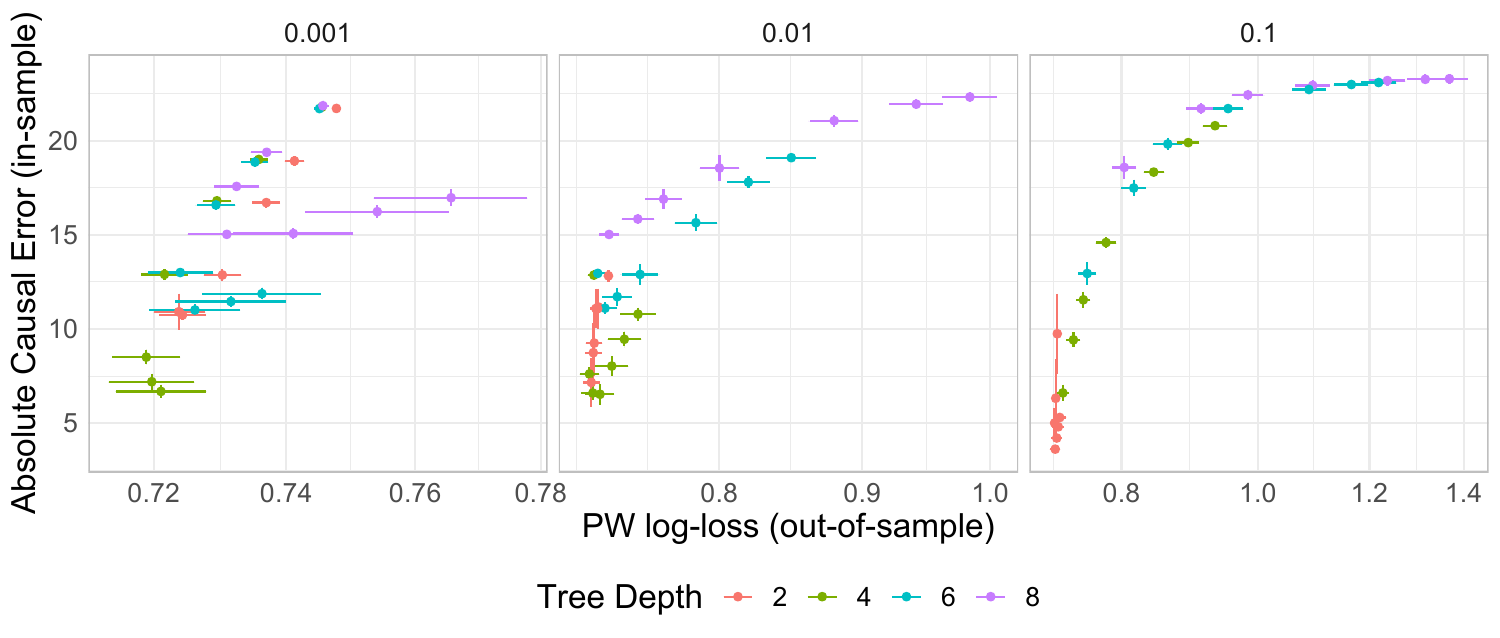}
    \caption{
    The estimated GBDT classifier error out-of-sample correlates strongly with the error of the causal estimate over a grid of hyperparameter values.
    The y-axis represents in-sample causal error, while the x-axis is the out-of-sample PW loss (i.e. out-of-sample imbalance).
    Hyperparameters tuned were the tree-depth of each decision tree (color), the learning rate, $\nu$ (columns) and the number of trees (not annotated, from 100 to 5000).
    }
    \label{fig:cv_results_oos}
\end{figure}

\begin{figure}
    \centering
    \includegraphics[width=\textwidth]{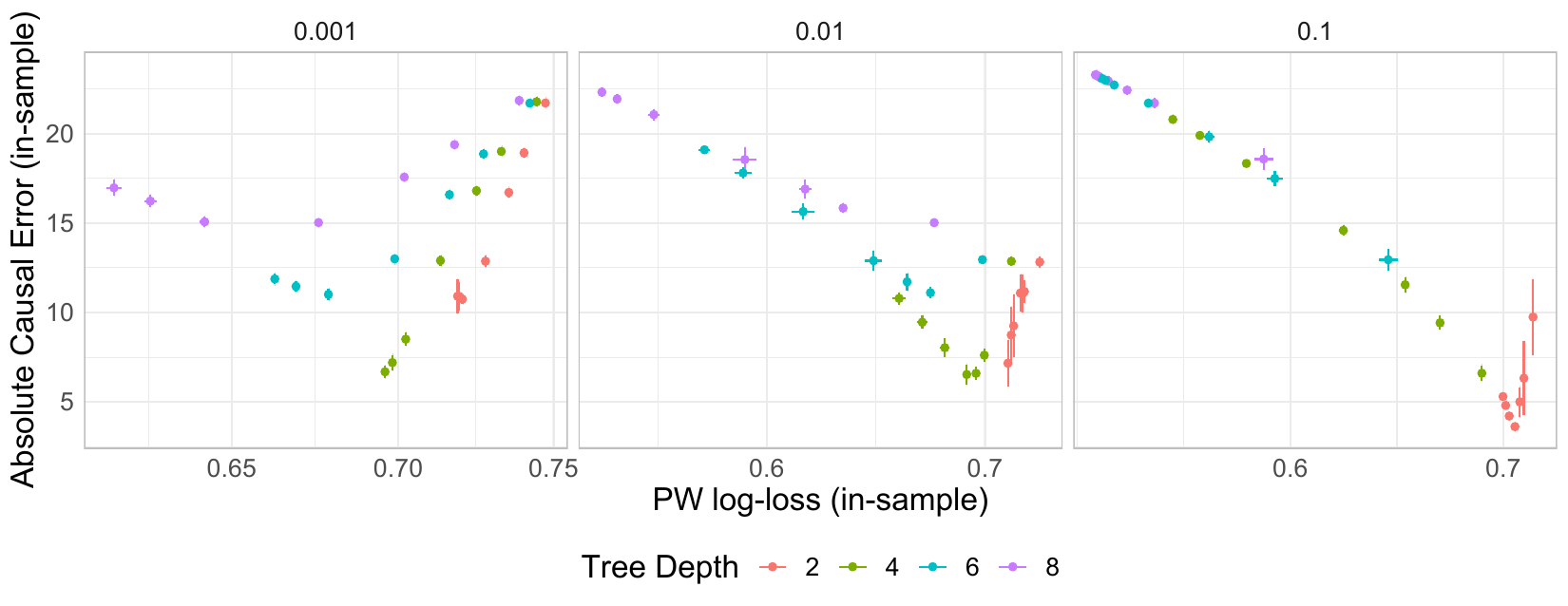}
    \caption{
    The estimated GBDT classifier error in-sample does not correlate with the error of the causal estimate over a grid of hyperparameter values.
    The y-axis represents in-sample causal error, while the x-axis is the in-sample PW loss (i.e. in-sample imbalance).
    Hyperparameters tuned were the tree-depth of each decision tree (color), the learning rate, $\nu$ (columns) and the number of trees (not annotated, from 100 to 5000).
    }
    \label{fig:cv_results_is}
\end{figure}

\begin{figure}
    \centering
    \includegraphics[width=.5\textwidth]{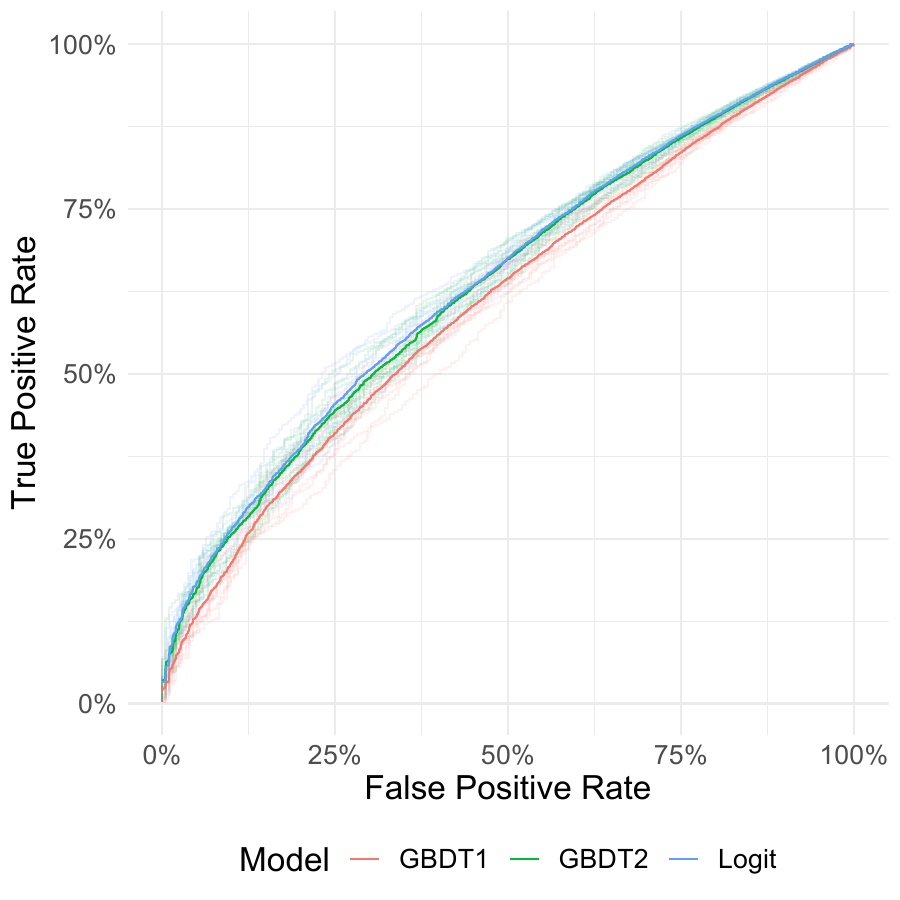}
    \caption{
    A ROC curve demonstrating that the linear model slightly out-performs the well-tuned GBDT across all false-positive rates.
    These two models have generally very similar performance, while the poorly tuned GBDT is substantially worse.
    The solid lines are the average over 10-fold cross-validation (the light lines in the background are each of the constituent folds).
    }
    \label{fig:cv_roc}
\end{figure}

In this section, we demonstrate how cross-validation may be used to effectively tune the permutation weight model.
For this experiment, we took one instance of the Kang-Schafer simulation (with a correctly specified linear model) described in section~\ref{subsec:kangschafer} with a sample size of $2000$ and performed 10-fold cross-validation on this data, measuring both in and out-of-sample errors.
Presented in figures~\ref{fig:cv_results_oos} and~\ref{fig:cv_results_is} are the results of this exercise for out-of-sample PW error and in-sample PW error, respectively.
Figure~\ref{fig:cv_roc} shows the ROC curve for three models: a well-tuned GBDT, a poorly tuned GBDT and a logistic regression.
For this linear data generating process, a poorly tuned GBDT is inadmissable no matter which proper scoring rule is used.
On the other hand, either the linear model or the well-tuned GBDT would be an acceptable choice depending on the choice of scoring rule.
Practitioners in this case would likely prefer to use the more parsimonious and interpretable linear model.

In the traditional causal inference environment, practitioners care about the in-sample causal error, rather than generalization error to other potential samples.
Minimizing classifier error, through proposition~\ref{prop:balance}, minimizes imbalance.
Minimization of in-sample imbalance may seem desirable since any residual imbalance can lead to bias in a purely weighting estimator.
Our results are shown in the first column of figures~\ref{fig:cv_results_oos} and~\ref{fig:cv_results_is}.
The primary takeaway from these results is that generalization performance of weights \emph{does} matter.
Simply minimizing in-sample imbalance is not necessarily the way to best optimize estimation accuracy.
More important than in-sample imbalance is \emph{out-of-sample} imbalance, which can be consistently measured through the PW error.
We can see clearly that while improving the PW loss out-of-sample brings with it corresponding improvements in in-sample causal error, this is not true for in-sample PW loss.
This represents classic over-fitting to the sample.
Importantly, looking at the error of permutation-weighting out-of-sample gives a reliable way to assess the quality of fit: choosing a permutation weighting model which generalizes well will, in turn, ensure that in-sample causal error is minimized.
In short, permutation weighting provides a reliable framework through which to use cross-validation for model selection and hyperparameter tuning for weighting.

\section{Conclusion}
Weighting is one of the most commonly applied estimators for causal inference. 
This work provides a new lens on weighting by framing the problem in terms of importance sampling towards the distribution of treatment and covariates that would be observed under an RCT.
Through this lens we introduced permutation weighting, which casts the balancing weights problem into generic binary classification and allows the standard machine learning toolkit to be applied to the problem.
We show that regret in this classification problem bounds the bias and variance of the causal estimation problem.
Thus, methods for regularization and model selection from the supervised learning literature can be used directly to manage the bias-variance tradeoff of this causal effect estimation problem.
Permutation weighting generalizes existing balancing schemes, admits selection via cross-validation, and provides a framework to sensibly integrate generic treatment types within the same weighting estimator.
Simulations show that permutation weighting outperforms existing estimation methods even in conditions unfavorable to the assumptions underlying the model.

\bibliographystyle{unsrtnat}
\bibliography{references}

\clearpage
\clearpage
\appendix\appendixpage

\section{Additional definitions}

Before presenting our results, we first restate the definition of Bregman divergences, which we denote with $B_g$.
Throughout, the specific form of $B_g$ is a consequence of the choice of classifier.

\begin{definition}[Bregman divergence~\citep{bregman1967relaxation}]
Define the Bregman generator, $g : S \to \mathbb{R}$, to be a convex, differentiable function.
The difference between the value of $g$ at point $s$ and the value of the first-order Taylor expansion of $g$ around point $s_0$ evaluated at point $s$ is given by
$B_g(s, s_0) \equiv g(s) - g(s_0) - \langle s - s_0, \nabla g(s_0) \rangle$.
\end{definition}

\section{Proofs of Propositions}

We begin by bounding the difference between the estimated and true density ratio weights, $|w(a, \mathbf{x}) - \hat{w}(a, \mathbf{x})|$ in terms of the classifier regret by using a Pinsker-type inequality.
This result will be used throughout in our technical proofs.

\begin{lem}
For any $(a, \mathbf{x})$, we have that $|w(a, \mathbf{x}) - \hat{w}(a, \mathbf{x})| \le
\frac{2}{\sqrt{{g''(1)}}} \sqrt{\operatorname{reg}(\hat{\eta} ; \mathcal{D}, \lambda)}$.
\label{lemma:pinsker1}
\end{lem}

\begin{proof}

Following Appendix B of \citet{menon2016linking}, let $\ubar{L} : [0, 1] \rightarrow \mathbb{R}$ be a concave differentiable function which provides the conditional Bayes risk for the classifier loss, i.e., 
\begin{align*}
    \lambda_{-1}(p)=\underline{\mathrm{L}}(p)-p \cdot \underline{\mathrm{L}}^{\prime}(p) \text { and } \lambda_{1}(p)=\underline{\mathrm{L}}(p)+(1-p) \cdot \underline{\mathrm{L}}^{\prime}(p)
\end{align*}
Recall, that the risk for the classifier $\hat{\eta}$ under loss $\lambda$ is then
\begin{align*}
2 \cdot \mathbb{L}(\hat{\eta} ; \mathcal{D}, \lambda) &=\mathbb{E}_{P}\left[\lambda_{1}(\hat{\eta}(a, \mathbf{x}))\right]+\mathbb{E}_{Q}\left[\lambda_{-1}(\hat{\eta}(a, \mathbf{x}))\right] \\
&=\mathbb{E}_{Q}\left[w(a, \mathbf{x}) \cdot \lambda_{1}(\hat{\eta}(a, \mathbf{x}))+\lambda_{-1}(\hat{\eta}(a, \mathbf{x}))\right]
\end{align*}
Let $g(z) = -(1 + z) \cdot \ubar{L}\left(\frac{z}{1 + z}\right)$.
Since $\hat{w} = \frac{\hat{\eta}}{1 - \hat{\eta}}$, 
\begin{align*}
g'(\hat{w}) = -(1 - \hat{\eta})\cdot \underline{\mathrm{L}}'(\hat{\eta}) -\underline{\mathrm{L}}(\hat{\eta})
= -\lambda_1(\hat{\eta})\\
\hat{w}\cdot g'(\hat{w}) - g(\hat{w}) = \lambda_{-1}(\hat{\eta})
\end{align*}
Therefore, we can write
\begin{align*}
    2 \cdot \mathbb{L}(\hat{\eta} ; \mathcal{D}, \lambda)=\mathbb{E}_{Q}\left[-w \cdot g^{\prime}(w)+\hat{w} \cdot g^{\prime}(\hat{w})-g(\hat{w})\right]
\end{align*}
Then, by \citet{reid2011information}, the Bayes risk for $\lambda$ is then given as
\begin{align*}
2 \cdot \mathbb{L}^{*}(\mathcal{D}, \lambda) 
&= 2 \cdot \min _{\hat{\eta}}{\mathbb{L}}(\hat{\eta} ; \mathcal{D}, \lambda) \\
&=-I_{g}(P, Q) \\
&=\mathbb{E}_{Q}[-g(w)],
\end{align*}
where $I_g$ is an $f$-divergence with generator $g$. Thus, the regret can be written as
\begin{align}
2 \cdot \operatorname{reg}(\hat{\eta} ; \mathcal{D}, \lambda) &=2 \cdot \mathbb{L}(\hat{\eta} ; \mathcal{D}, \lambda)-2 \cdot \mathbb{L}^{*}(\mathcal{D}, \lambda) \nonumber \\
&=\mathbb{E}_{Q}\left[-w \cdot g^{\prime}(w)+\hat{w} \nonumber \cdot g^{\prime}(\hat{w})-g(\hat{w})+g(w)\right] \nonumber \\
&=\mathbb{E}_{Q}\left[B_{g}(w, \hat{w})\right] 
\label{eq:regret}
\end{align}
This, in turn, implies that the $f$-divergence, $I_g$ can be written as
\begin{align}
\label{eq:bregF}
    I_g(P, Q) 
    &= -2 \cdot \mathbb{L}^{*}(\mathcal{D}, \lambda) 
    = \mathbb{E}_{Q}\left[B_{g}(w, \hat{w})\right] - 2 \cdot \mathbb{L}(\hat{\eta} ; \mathcal{D}, \lambda)
\end{align}
Under additional assumptions on $g$, the following theorem, due to \citet{gilardoni2008pinsker}, provides a bound in terms of the total variation distance, $V := V(P, Q)$. 
\begin{thm}[Pinsker type inequality for $f$-divergences.]
\label{thm:genPinsker}
Suppose that the convex function $g$ is differentiable up to order 3 at $u=1$ with $g^{\prime \prime}(1)>0$. Let $D_g$ denote the $f$-divergence generated by $g$. Then, $D_{g} \geq \frac{g^{\prime \prime}(1)}{2} V^{2} .$ The constant $\frac{g^{\prime \prime}(1)}{2}$ is best possible.
\end{thm} 
Applying Theorem \ref{thm:genPinsker} to equation~\eqref{eq:bregF} gives
\begin{align*}
    I_g(P, Q)  = \mathbb{E}_{Q}\left[B_{g}(w, \hat{w})\right] - 2 \cdot \mathbb{L}(\hat{\eta} ; \mathcal{D}, \lambda) \geq \frac{g''(1)}{2}V^2.
\end{align*}
Therefore.
\begin{align*}
    V &\leq \sqrt{\frac{2\left(\mathbb{E}_{Q}\left[B_{g}(w, \hat{w})\right] - 2 \cdot \mathbb{L}(\hat{\eta} ; \mathcal{D}, \lambda)\right)}{g''(1)}} \\
    &\leq \sqrt{\frac{2}{g''(1)} \mathbb{E}_{Q}\left[B_{g}(w, \hat{w})\right]}.
\end{align*}
This can be related directly to the classification problem considered in this paper with an application of equation~\eqref{eq:regret} as shown above, yielding
\begin{align*}
    V \leq \frac{2}{\sqrt{{g''(1)}}} \sqrt{\operatorname{reg}(\hat{\eta} ; \mathcal{D}, \lambda)}
\end{align*}
Then, by definition of the total variation distance, this inequality holds for any $|w(a, \mathbf{x}) - \hat{w}(a, \mathbf{x})|$. 
\end{proof}

In the table below, we verify the conditions of the generator $g$ required for \ref{thm:genPinsker} corresponding to different $f$-divergences:
\begin{table}[]
\centering
\begin{tabular}{llll}
\hline
Divergence                & $g(u)$ & $g''(1)$ & $g'''(1)$ \\ \hline
Kullback-Leibler          &  $u\log(u)$    &   1     & -1        \\
Triangular Discrimination & $\frac{(u - 1)^2}{u+1}$     &  1      & -3/2  \\
Jensen-Shannon            & $\frac{u}{2} \log\frac{u}{u + 1} - \frac{1}{2} \log\frac{u + 1}{4}$     &    1/4    &    -3/8     \\
Jeffreys                  &  $(u-1)\log(u)$    &     2   &  -3       \\
Hellinger                 &   $(\sqrt{u} - 1)^2$   &    1/2    &    -3/4     \\
Pearson $\mathcal{X}^2$               &  $(u - 1)^2$    &    2    &  0  
\end{tabular}
\end{table}

\pwbias*
\begin{proof}
Let $w(a_i, \mathbf{x}_i) = \frac{p(a_i) p(\mathbf{x}_i)}{p(a_i, \mathbf{x}_i)}$ and $\hat{w}(a(a_i, \mathbf{x}_i)$ be the empirical estimate of $w(a_i, \mathbf{x}_i)$.
\begin{align*}
    \mathbb{E}_P[y\hat{w}(a, \mathbf{x})] 
    &= \mathbb{E}_P\left[ y_i(w(a_i, \mathbf{x}_i) + (\hat{w}(a_i, \mathbf{x}_i)- w(a_i, \mathbf{x}_i)))\right]\\
    &= \mathbb{E}_P[y_i w(a_i,\mathbf{x}_i)] + \mathbb{E}_P \left[ y_i(\hat{w}(a_i, \mathbf{x}_i)- w(a_i, \mathbf{x}_i)) \right] \\
    &= \mathbb{E}_Q[y_i] + \mathbb{E}_P \left[ y_i(\hat{w}(a_i, \mathbf{x}_i)- w(a_i, \mathbf{x}_i)) \right]
\end{align*}
By Lemma \ref{lemma:pinsker1}, the bias can then be written as
\begin{align*}
    \mathbb{E}_P \left[ y_i(\hat{w}(a_i, \mathbf{x}_i)- w(a_i, \mathbf{x}_i)) \right]
    &\leq \mathbb{E}_P \left[ |y_i| |\hat{w}(a_i, \mathbf{x}_i)- w(a_i, \mathbf{x}_i| \right] \\
    &\leq \mathbb{E}_P \left[ |y_i| \frac{2}{\sqrt{{g''(1)}}} \sqrt{\operatorname{reg}(\hat{\eta} ; \mathcal{D}, \lambda)} \right]
\end{align*}

\end{proof}

\pwvariance*
\begin{proof}
The proofs follows via Lemma \ref{lemma:pinsker1}.
We first note that a trivial upper bound for the variance is given by the second moment.
We consider the second moment of the estimator given as 
\begin{align*}
\mathbb{E}_P\left[ \left(y\hat{w}\right)^2\right]
&= \mathbb{E}_P \left[ y^2\hat{w}^2 \right] \\
&= \mathbb{E}_P\left[y^2(w + \left(\hat{w} - w\right) )^2\right]\\
&= \mathbb{E}_P\left[y^2 w^2\right] + \mathbb{E}_P\left[y^2\left(2{w}\left(\hat{w} - w\right) + \left(\hat{w} - w\right)^2\right)
\right]\\
&\leq \mathbb{E}_P\left[y^2 w^2\right] + \mathbb{E}_P\left[y^2\left(\frac{4 w}{\sqrt{{g''(1)}}} \sqrt{\operatorname{reg}(\hat{\eta} ; \mathcal{D}, \lambda)} + \frac{4}{g''(1)} \operatorname{reg}(\hat{\eta} ; \mathcal{D}, \lambda) \right) \right] \\
&= \mathbb{E}_P\left[y^2 w^2\right] +
\frac{4}{\sqrt{{g''(1)}}} \sqrt{\operatorname{reg}(\hat{\eta} ; \mathcal{D}, \lambda)} 
\mathbb{E}_P\left[y^2\left(w + \frac{1}{\sqrt{g''(1)}} \sqrt{\operatorname{reg}(\hat{\eta} ; \mathcal{D}, \lambda)} \right) \right]
\end{align*}

\end{proof}

\pwconsistency*
\begin{proof}
From Propositions~\ref{prop:bregmanBoundedBias} and ~\ref{prop:bregmanBoundedVariance}, by minimizing an appropriate classifier loss, we can bound the bias and variance in terms of the classifier regret. 
Thus, under Assumption \ref{assump:consist}, and given bounded $y$, the bias and variance tend to 0 as $n \longrightarrow \infty$.
\end{proof}
\pwbalance*
\begin{proof}
We will focus on the case of attaining balance between a source distribution, $P$ and a target distribution $Q$ defined over a common support. 
Define $w(a_i, x_i) = \frac{q(a_i, x_i)}{p(a_i, x_i)}$ and $\hat{w}(a_i, x_i)$ be an estimate of $w$.
\begin{align*}
&\left\|\mathbb{E}_{P}\left[{\phi(a_i)\otimes\psi(x_i)}{\hat{w}(a_i, x_i)}\right] - \mathbb{E}_{Q}\left[\phi(a_i)\otimes\psi(x_i)\right]\right\|_p\\
=&\left\|\mathbb{E}_{P}\left[{\phi(a_i)\otimes\psi(x_i)}{\hat{w}(a_i, x_i)}\right] - \mathbb{E}_{P}\left[\phi(a_i)\otimes\psi(x_i)\frac{q(a_i, x_i)}{p(a_i, x_i)}\right]\right\|_p\\    
=&\left\|\sum_i^N {\phi(a_i)\otimes\psi(x_i)}{\hat{w}(a_i, x_i)} p(a_i, x_i) - \sum_i^N \phi(a_i)\otimes\psi(x_i)\frac{q(a_i, x_i)}{p(a_i, x_i)}p(a_i, x_i)\right\|_p\\
=&\left\|\sum_i^N \phi(a_i)\otimes\psi(x_i){p(a_i, x_i)}{\hat{w}(a_i, x_i)}  - \phi(a_i)\otimes\psi(x_i)q(a_i, x_i)\right\|_p\\
=&\left\|\sum_i^N \phi(a_i)\otimes\psi(x_i){p(a_i, x_i)}{\left(w(a_i, x_i) + (\hat{w}(a_i, x_i) - w(a_i, x_i))\right)}  - \phi(a_i)\otimes\psi(x_i)q(a_i, x_i)\right\|_p\\
=&\left\|\sum_i^N \phi(a_i)\otimes\psi(x_i){\left(p(a_i, x_i)w(a_i, x_i) + p(a_i, x_i)(\hat{w}(a_i, x_i) - w(a_i, x_i))\right)}  - \phi(a_i)\otimes\psi(x_i)q(a_i, x_i)\right\|_p\\
=&\left\|\sum_i^N \phi(a_i)\otimes\psi(x_i)q(a_i, x_i) + \phi(a_i)\otimes\psi(x_i)p(a_i, x_i)(\hat{w}(a_i, x_i) - w(a_i, x_i)) - \phi(a_i)\otimes\psi(x_i)q(a_i, x_i)\right\|_p\\
=&\left\|\sum_i^N  \phi(a_i)\otimes\psi(x_i)p(a_i, x_i)(\hat{w}(a_i, x_i) - w(a_i, x_i)) \right\|_p\\
=&\left\|\mathbb{E}_{P}\left[\phi(a_i)\otimes\psi(x_i)(\hat{w}(a_i, x_i) - w(a_i, x_i))\right]\right\|_p
\end{align*}
Then, by Lemma \ref{lemma:pinsker1}, the result follows.
\end{proof}

\section{Alternative Derivation of the Balance Conditions of Permutation Weighting with Logistic Regression}
\label{app:pw_as_logit}
Given the prominence of mean balance within the weighting literature, we now provide additional focus on the mean balance condition of permutation weighting.
Specifically, we derive the first order score condition for to allow for optimization via the  generalized method of moments for permutation weighting using logistic regression and a feature specification of $c \sim \text{logit}(\beta_0 + \beta_1 a_i \mathbf{x_i} + \beta_2 x_i)$, where $\beta$ are the regression coefficients.  
We define the model as $\pi_\beta(a, X)$ as a function that infers the probability that an instance, $a_i, X_i$ belongs to the resampled data, denoted $C=1$ or the original observed data, denoted as $C=0$. 
In other words $\pi_\beta(a_i, X_i) = p(C_i | a_i, X_i)$.
We assume that there are $N$ total data instances, with $\frac{1}{2}$ of the instances being the resampled data and $\frac{1}{2}$ of the instances being the original data.
Note that $\pi_\beta(a_i, X_i)$ is a convex function and that its parameters $\beta$, denote as $\pi'_\beta(a, X)$ is twice continuously differentiable as a consequence of using logistic loss~(all strictly proper scoring rules obey these properties).
We assume that we wish to maximize the log-likelihood of $\pi_\beta$

\begin{align*}
    &\Hat{\beta}_{MLE} = \arg\max_{\beta \in \Theta} \sum_i^N C_i\log\left(\pi_\beta(X_i, a_i)\right) + (1 - C_i)\log\left(1 - \pi_\beta(X_i, a_i)\right)
\end{align*}

which gives the following first order condition

\begin{align*}
    &\sum_i^N s_\beta(C_i, a_i, X_i) = 0\\
    &s_\beta(C_i, a_i, X_i) = \sum_i^N\frac{C_i\pi'_\beta(a_i, X_i)}{\pi_\beta(a_i, X_i)} - \frac{(1 - C_i)\pi'_\beta(a_i, X_i)}{1 - \pi_\beta(a_i, X_i)}\\
    &= \sum_i^N\frac{C_i\pi'_\beta(a_i, X_i)}{\pi_\beta(a_i, X_i)} - \sum_i^N\frac{(1 - C_i)\pi'_\beta(a_i, X_i)}{1 - \pi_\beta(a_i, X_i)}\\ 
\end{align*}

Letting $\tilde{\mathbf{x}} = \left[\mathbf{x}, a\right]$

\begin{align}
\pi' = \frac{\exp(\theta_{a, \mathbf{x}} a_i\mathbf{x}_i + \theta_{\mathbf{x}_i}\mathbf{x}_i)}{\left(1 + \exp(\theta_{a, \mathbf{x}} a_i\mathbf{x}_i + \theta_{\mathbf{x}_i}\mathbf{x}_i)\right)^2}\tilde{\mathbf{x}}_i
\end{align}

The left-hand side may be rewritten as

\begin{align*}
    &= \sum_i^N\frac{C_i\pi'_\beta(a_i, X_i)}{\pi_\beta(a_i, X_i)}
    = \sum_i^N\frac{C_i \frac{\exp(\theta_{a, \mathbf{x}} a_i\mathbf{x}_i + \theta_{\mathbf{x}_i}\mathbf{x}_i)}{\left(1 + \exp(\theta_{a, \mathbf{x}} a_i\mathbf{x}_i + \theta_{\mathbf{x}_i}\mathbf{x}_i)\right)^2}\tilde{\mathbf{x}}_i}{\frac{\exp(\theta_{a, \mathbf{x}} a_i\mathbf{x}_i + \theta_{\mathbf{x}_i}\mathbf{x}_i)}{\left(1 + \exp(\theta_{a, \mathbf{x}} a_i\mathbf{x}_i + \theta_{\mathbf{x}_i}\mathbf{x}_i)\right)}}
    = \sum_i^N\frac{C_i \tilde{\mathbf{x}}_i}{{\left(1 + \exp(\theta_{a, \mathbf{x}} a_i\mathbf{x}_i + \theta_{\mathbf{x}_i}\mathbf{x}_i)\right)}}
\end{align*}

The right-hand side may be rewritten as:

\begin{align*}
    &\sum_i^N\frac{(1 - C_i)\pi'_\beta(a_i, X_i)}{1 - \pi_\beta(a_i, X_i)} 
    = \sum_i^N\frac{(1 - C_i) \frac{\exp(\theta_{a, \mathbf{x}} a_i\mathbf{x}_i + \theta_{\mathbf{x}_i}\mathbf{x}_i)}{\left(1 + \exp(\theta_{a, \mathbf{x}} a_i\mathbf{x}_i + \theta_{\mathbf{x}_i}\mathbf{x}_i)\right)^2}\tilde{\mathbf{x}}_i}{1 - \frac{\exp(\theta_{a, \mathbf{x}} a_i\mathbf{x}_i + \theta_{\mathbf{x}_i}\mathbf{x}_i)}{\left(1 + \exp(\theta_{a, \mathbf{x}} a_i\mathbf{x}_i + \theta_{\mathbf{x}_i}\mathbf{x}_i)\right)}}\\
    &= \sum_i^N\frac{( 1 - C_i) \frac{\exp(\theta_{a, \mathbf{x}} a_i\mathbf{x}_i + \theta_{\mathbf{x}_i}\mathbf{x}_i)}{\left(1 + \exp(\theta_{a, \mathbf{x}} a_i\mathbf{x}_i + \theta_{\mathbf{x}_i}\mathbf{x}_i)\right)^2}\tilde{\mathbf{x}}_i}{\frac{1}{\left(1 + \exp(\theta_{a, \mathbf{x}} a_i\mathbf{x}_i + \theta_{\mathbf{x}_i}\mathbf{x}_i)\right)}} 
    = \sum_i^N (1 - C_i) \frac{\exp(\theta_{a, \mathbf{x}} a_i\mathbf{x}_i + \theta_{\mathbf{x}_i}\mathbf{x}_i)}{\left(1 + \exp(\theta_{a, \mathbf{x}} a_i\mathbf{x}_i + \theta_{\mathbf{x}_i}\mathbf{x}_i)\right)}\tilde{\mathbf{x}}_i\\
    & = \sum_i^N \frac{(1 - C_i) \tilde{\mathbf{x}}_i}{\left(1 + \exp(-\theta_{a, \mathbf{x}} a_i\mathbf{x}_i - \theta_{\mathbf{x}_i}\mathbf{x}_i)\right)}
\end{align*}

Substituting both in we have the following score condition: 

\begin{align}
\label{eq:linearBalance}
&0 = \sum_i^N\frac{C_i \tilde{\mathbf{x}}_i}{{\left(1 + \exp(\theta_{a, \mathbf{x}} a_i\mathbf{x}_i + \theta_{\mathbf{x}_i}\mathbf{x}_i)\right)}} - 
    \sum_i^N \frac{(1 - C_i) \tilde{\mathbf{x}}_i}{\left(1 + \exp(-\theta_{a, \mathbf{x}} a_i\mathbf{x}_i - \theta_{\mathbf{x}_i}\mathbf{x}_i)\right)}\\
\nonumber&= \sum_i^N\frac{C_i a_i\tilde{\mathbf{x}}_i}{{\left(1 + \exp(\theta_{a, \mathbf{x}} a_i\mathbf{x}_i + \theta_{\mathbf{x}_i}\mathbf{x}_i)\right)}} + \frac{C_i (1-a)\tilde{\mathbf{x}}_i}{{\left(1 + \theta_{\mathbf{x}_i}\mathbf{x}_i)\right)}}- 
    \frac{(1 - C_i) a_i \tilde{\mathbf{x}}_i}{\left(1 + \exp(-\theta_{a, \mathbf{x}} a_i\mathbf{x}_i - \theta_{\mathbf{x}_i}\mathbf{x}_i)\right)} + \frac{(1 - C_i) (1 - a_i) \tilde{\mathbf{x}}_i}{\left(1 + \exp(-\theta_{\mathbf{x}_i}\mathbf{x}_i)\right)}\\
\nonumber&= \sum_i^N
\left(
\frac{C_i a_i\tilde{\mathbf{x}}_i}{{\left(1 + \exp(\theta_{a, \mathbf{x}} a_i\mathbf{x}_i + \theta_{\mathbf{x}_i}\mathbf{x}_i)\right)}} - 
\frac{(1 - C_i) a_i \tilde{\mathbf{x}}_i}{\left(1 + \exp(-\theta_{a, \mathbf{x}} a_i\mathbf{x}_i - \theta_{\mathbf{x}_i}\mathbf{x}_i)\right)} \right)+ 
\left(\frac{C_i (1-a)\tilde{\mathbf{x}}_i}{{\left(1 + \exp(\theta_{\mathbf{x}_i}\mathbf{x}_i))\right)}} 
 -
\frac{(1 - C_i) (1 - a_i) \tilde{\mathbf{x}}_i}{\left(1 + \exp(-\theta_{\mathbf{x}_i}\mathbf{x}_i)\right)}\right)\\
\nonumber&= \sum_i^N
\left(
\frac{C_i a_i\tilde{\mathbf{x}}_i}{{\left(1 + \exp(\theta_{a, \mathbf{x}} a_i\mathbf{x}_i + \theta_{\mathbf{x}_i}\mathbf{x}_i)\right)}} - 
\frac{(1 - C_i) a_i \tilde{\mathbf{x}}_i}{\left(1 + \exp(-\theta_{a, \mathbf{x}} a_i\mathbf{x}_i - \theta_{\mathbf{x}_i}\mathbf{x}_i)\right)} \right)+ 
\left(\frac{C_i (1-a)\tilde{\mathbf{x}}_i}{{\exp\left(1 + \theta_{\mathbf{x}_i}\mathbf{x}_i)\right)}} 
 -
\frac{(1 - C_i) (1 - a_i) \tilde{\mathbf{x}}_i}{\left(1 + \exp(-\theta_{\mathbf{x}_i}\mathbf{x}_i)\right)}\right)
\end{align}

A interpretation of the final term in equation \ref{eq:linearBalance} is the difference in reweighted means between treatment and control and overall population. 
The case of balance, i.e., when both treatment and control are reweighted to the marginal mean of $\mathbf{X}$, satisfies the score condition.
\section{Connections to Existing Weighting Estimators}
\label{app:connection}
We now examine the connection between permutation and a number of covariate balancing estimators in the literature. 
In what follows, we first revisit the relationship with stabiized inverse propensity score weighting, look at score based estimates, i.e. estimators which can be estimated using the generalized method of moments with a particular focus on the covariate balancing propensity score~\cite{imai2014covariate}, then examine margin based estimators and their kernel extension, establishing an equivalence between permutation weighting and kernel mean matching~\cite{huang2007correcting}, kernel balancing~\cite{hazlett2016kernel}, and Kernel based covariate functional balancing~\cite{wong2017kernel} and relate permutation to stabilized balancing weights~\cite{zubizarreta2015stable}.

\subsection{Stabilized Inverse Propensity Score}
Perhaps the most immediately evident equivalence is to the stabilized inverse propensity score~(IPSW)~\cite{robins1997causal}. 
This relationship is addressed in the main text; it is included in this section for the benefit of completeness. 
Recall the definition of the stabilized propensity score weight is simply the marginal density of treatment divide by the conditional density of treatment given covariates, i.e., $\frac{p(a)}{p(a | \mathbf{x})}$.
Employing simply algebra we see that this quantity will be equivalent to permutation weights, i.e. $\frac{p(a)p(\mathbf{x})}{p(a, \mathbf{x})}$, under correct specification of the conditional density. 
The crucial difference between permutation weighting and IPSW comes under mis-specification.
Permutation weighting will still seek balance with respect to the conditions implied by the classifier.
IPSW, on the other hand, may fail to seek balance under mis-specification.
This can result in substantial bias, as we have seen in the empirical results of the main text. 

\subsection{Covariate Balancing Propensity Scores}
\label{app:cbps_connection}
We first note the score condition of the covariate balancing propensity score~\cite{imai2014covariate}: 

\begin{align}
0 = \sum_i^N\frac{a_i \mathbf{{x}}_i}{{\left(1 + \exp(-\theta_{\mathbf{x}_i}\mathbf{x}_i)\right)}} - 
    \sum_i^N \frac{(1 - a_i) \mathbf{{x}}_i}{\left(1 + \exp(\theta_{\mathbf{x}_i}\mathbf{x}_i)\right)}
\end{align}

Recalling the derivation of the score condition for PW with logistic loss from equation \ref{eq:linearBalance} provides a simple comparison: 

\begin{align*}
0 = \sum_i^N &
\left(
\frac{C_i a_i\tilde{\mathbf{x}}_i}{{\left(1 + \exp(\theta_{a, \mathbf{x}} a_i\mathbf{x}_i + \theta_{\mathbf{x}_i}\mathbf{x}_i)\right)}} - 
\frac{(1 - C_i) a_i \tilde{\mathbf{x}}_i}{\left(1 + \exp(-\theta_{a, \mathbf{x}} a_i\mathbf{x}_i - \theta_{\mathbf{x}_i}\mathbf{x}_i)\right)} \right)\\
&+ 
\left(\frac{C_i (1-a)\tilde{\mathbf{x}}_i}{{\exp\left(1 + \theta_{\mathbf{x}_i}\mathbf{x}_i)\right)}} 
 -
\frac{(1 - C_i) (1 - a_i) \tilde{\mathbf{x}}_i}{\left(1 + \exp(-\theta_{\mathbf{x}_i}\mathbf{x}_i)\right)}\right)
\end{align*}

Here we can see that both estimators are explicitly minimizing a balance condition, PW to the product, i.e. independent, distribution and CBPS between classes.
In large samples these are equivalent conditions.
However, an interesting interpretation emerges when we consider what happens in smaller samples. 
Permutation weighting will attempt to match the level of balance that exists in the empirical sample, which provides a data dependent regularization.
This regularization likely explains the improvement of PW over CBPS in the case of the synthetic experiments involving mis-specification.

We note that a similar derivation yields the following for inverse propensity score weighting: 

\begin{align}
0 = \sum_i^N\frac{a_i \mathbf{{x}}_i}{{\left(1 + \exp(\theta_{\mathbf{x}_i}\mathbf{x}_i)\right)}} - 
    \sum_i^N \frac{(1 - a_i) \mathbf{{x}}_i}{\left(1 + \exp(-\theta_{\mathbf{x}_i}\mathbf{x}_i)\right)}
\end{align}

Here we can see that balance is \textit{not} directly optimized for, which explains much of the poor performance of inverse propensity score weighting in the synthetic experiments with a misspecified estimator. 

\subsection{Stabilized Balancing Weights, Kernel Mean Matching, and Kernel Balancing}

We will now briefly introduce MMD, weighting methods predicated on MMD, e.g. \cite{huang2007correcting, gretton2009covariate}, followed by a discussion of their connection to stabilized balancing weights~\cite{zubizarreta2015stable}. 

The maximum mean discrepancy~(MMD)~\cite{gretton2012kernel}, is a two-sample test that distinguishes between two candidate distributions by finding the maximum mean distance between the means of the two samples after transformation, i.e., 

\begin{align}
    \label{eq:mmd} \sup_{f\in\mathcal{F}} \left( \mathbb{E}_{a\sim A}\left[f(a)\right] - \mathbb{E}_{b\sim B}\left[f(b)\right] \right)
\end{align}

When $\mathcal{F}$ is a reproducing kernel Hilbert space this can be estimated as the squared difference of their means in feature space. 
Letting $\phi(\cdot)$ be a kernel associated with a random variable $A$ and $\psi(\cdot)$ be the kernel associated with random variable $B$, the finite sample estimate of equation \ref{eq:mmd} is given as

\begin{align*}
  \text{MMD}(A, B) =  \left\|  \frac{1}{N}\sum_i^N \phi(a_i) -  \frac{1}{M}\sum_j^M \psi(b_j)\right\|^2
\end{align*}

with $N$ and $M$ being the size of the samples drawn from $A$ and $B$, respectively. 
The value of MMD reflects the maximum distance between these means 
There are a couple of points worth noting. 
First, if the kernel being used obeys certain properties, i.e is characteristic~\cite{sriperumbudur2008injective}, the MMD is able to differentiate between two exponential-family distributions on an arbitrary number of moments~\cite{gretton2012kernel}. 
Second, when a linear kernel is employed this is value is simply the squared difference in means between the two groups. 

MMD has been used throughout the literature as an objective for minimizing imbalance. 
Within the context of domain adaptation, \cite{huang2007correcting} introduce kernel mean matching~(KMM) which
defines an optimization procedure that seeks to find a set of weights such that the distance between the target and source distribution is minimized, specifically

\begin{align}
    &\label{eq:KMM}\min_\beta \left\| \frac{1}{N}\sum_i^N \beta(a_i)\phi(a_i) - \frac{1}{M}\sum_j \psi(b_j) \right\|^2\\
    &\nonumber\text{Such that } \beta(a) > 0, \sum_i^N \beta(a_i) = 1
\end{align}

This procedure was later rediscovered for the task of balancing weights by \cite{hazlett2016kernel}. 
Somewhat surprisingly, the connection to permutation weighting can be easily obtained via results currently found in the literature. 
\cite{reid2011information} relate a pessimistic MMD to the support vector machine~(SVM), seeking to maximize the MMD by solving the SVM problem

\begin{align*}
    &\min_\alpha \sum_i^M \sum_j^M \alpha_i\alpha_j c_i c_j k(x_i, x_j)\\
    &\text{Such that } \alpha \geq 0\\
    &\sum_i^m \alpha_i y_i = \frac{m^+ - m^-}{m}\\
    &\sum_i^m \alpha_i = 1
\end{align*}

where $c$ indicates which dataset a sample has been drawn from and is coded $\{-1, 1\}$.
\cite{bickel2007discriminative}~(Section 8) shows that solving the KMM objective is equivalent to solving the above SVM problem under the additional modification that the values of $\alpha$ for are fixed to a constant for one class, producing a Rocchio-style approximation~\cite{joachims1997probabilistic} to the SVM. 
In both cases the final weighting is given directly by taking the value of the dual weights~($\alpha$).

The aforementioned classifiers may be employed within the context of permutation weighting by considering the two samples to be the observed data with a target distribution of the resampled data~(as we have assumed throughout).
In order to use the dual weights directly the bootstrap procedure is replaced by an average over permutations. 
Alternatively the weight function may be used directly by considering $\exp(w(\mathbf{x_i}, a_i)$ as in \cite{bickel2007discriminative}. 
The benefit of the latter approach is the ability to use cross validation for setting the hyper-parameters of the classifiers. 
Asymptotically, as the independence property is obeyed by the resampled data, permutation weighting and these procedures are equivalent in the binary treatment setting.
To see why this is the case, consider the explicit form of permutation weighting under MMD loss:

\begin{align*}
    \sup_{f\in \mathcal{F}} \left(\mathbb{E}_{a, x \sim p(A)p(\mathbf{X})}\left[f(a, x)\right] - \mathbb{E}_{a, x \sim p(A, \mathbf{X})} \left[f(a, x)\right] \right)
\end{align*}

where we have again assumed that $\mathcal{F}$ is a reproducing kernel Hilbert space. 
This is precisely the Hilbert-Schmidt independence criterion~\cite{GreHerSmoBouSch05}, which was shown by \cite{song2008learning} to be equivalent to the maximum mean discrepancy between the two $\mathbf{X}$ samples associated with treatment and control, respectively, when $A$ is binary.
In the non-binary case the permutation weighting setup defines a novel kernel balancing estimator for general treatments.

Finally, we examine the relationship between permutation weighting and the stable balancing weights of \citet{zubizarreta2015stable}.
\citet{zubizarreta2015stable} defined the following quadratic program to infer what he refers to as stable balancing weights: 

\begin{align*}
    &\min_w \left\|\mathbf{w} - \overline{\mathbf{w}}\right\|^2_2\\
    &\text{Subject to}\\
    &\left| \mathbf{w}^T X_{\text{control}_p} - \overline{\mathbf{X}}_{\text{test}_p} \right| < \delta, p = 1, \dots, k\\
    &\sum w = 1\\
    &w \geq 0\\
\end{align*}

Intuitively this attempts to minimize the variance of the weights subject to constraints on marginal balance conditions. 
Comparing this to the kernel mean matching problem~(equation \ref{eq:KMM}), we see that stable balancing weights emphasize uniform weights subject to a constraint of predetermined levels of marginal balance. 
Kernel mean matching on the other hand, seeks to minimize the maximum discrepancy between the two distributions. 
Minimizing the discrepancy rather than setting it to a fixed level is that it removes a large amount of possible human induced error in the form of additional hyperparameters.
While the MMD approach does not have an explicit mechanism to reduce variance, an approximation can be applied by solving the SVM problem using a $\nu$-SVM~\cite{scholkopf2001learning} which imposes an additional constraints that limits the size of individual weights.

\section{Continuous Kang-Schafer Data Generating Process}
\label{app:kangschafer-continuous}

The same basic setup can be extended to the continuous treatment case by replacing the Bernoulli treatment assignment with a continuous analogue.
For the following simulations, we simulate treatment dosage as a linear function as in \citep{kang2007demystifying}, but adding in standard normal noise.
Finally, the dosage enters the outcome model through a logit function to introduce a small non-linearity in response to dose.

\begin{align*}
X_k &= \mathcal{N}(0,1) \quad \forall k \in \{1,2,3,4\}\\
\epsilon &\sim \mathcal{N}(0,1)\\
A\mid X &= X_1 - 0.5 X_2 + 0.25 X_3 + 0.1 X_4 + \epsilon \\
\E[Y\mid A, X] &= 210 + \text{logit}(A) + 27.4 X_1 + 13.7 X_2 + 13.7 X_3 + 13.7 X_4\\
Y\mid A, X &= \mathcal{N}(\E[Y\mid A, X], 1)
\end{align*}

Misspecification is handled identically to the binary case.

\begin{figure}
    \centering
    \includegraphics[width=.8\textwidth]{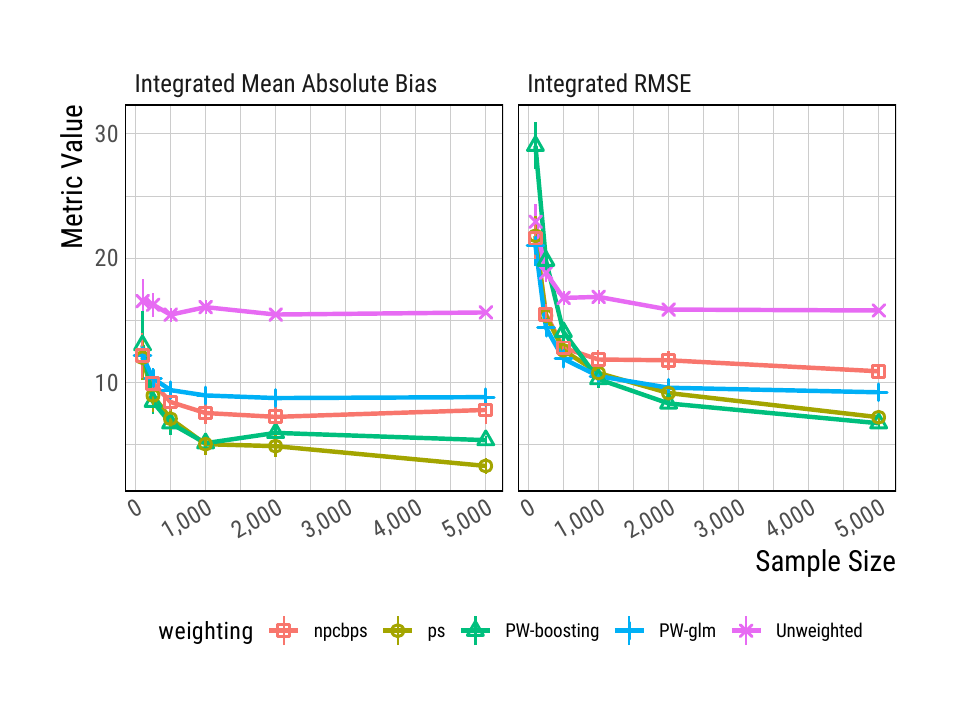}
    \caption{
    Continuous Kang-Schafer simulation under correct specification of confounding variables.
    \texttt{npcbps} is non-parametric covariate balancing propensity scores \citep{fong2018covariate}.
    \texttt{ps} is a normal-linear regression based propensity score model.
    \texttt{PW-boosting} is a permutation weighting model using a gradient boosted decision tree.
    \texttt{PW-glm} is a permutation weighting model using a logistic regression.
    \texttt{Unweighted} uses no weighting.
    }
    \label{fig:KS-cont-correct}
\end{figure}

\begin{figure}
    \centering
    \includegraphics[width=.8\textwidth]{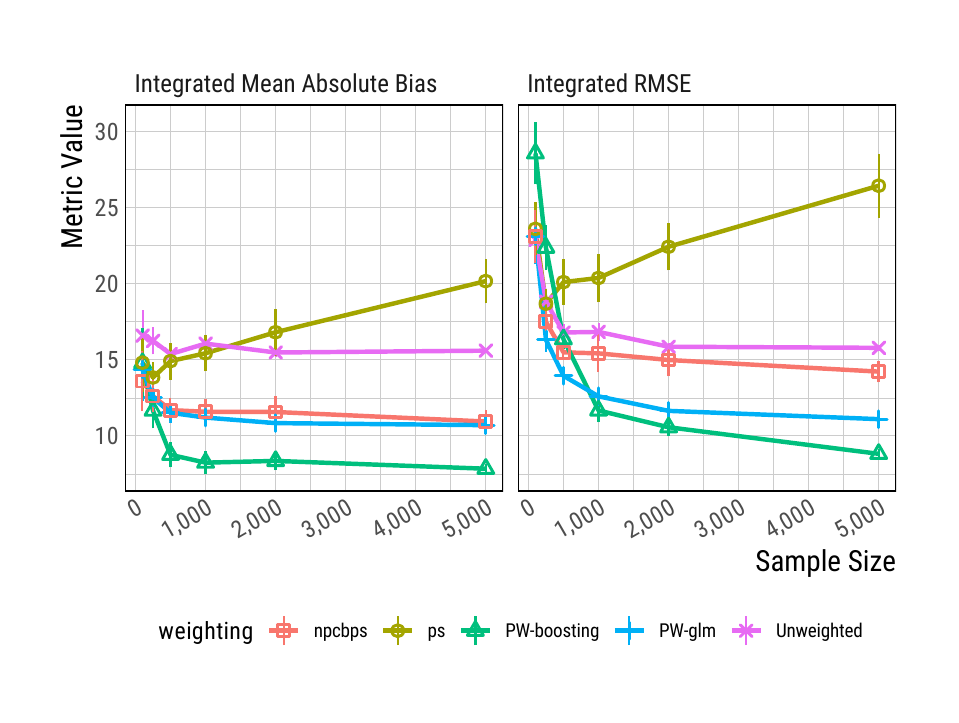}
    \caption{
    Continuous Kang-Schafer simulation under misspecification of confounding variables.
    \texttt{npcbps} is non-parametric covariate balancing propensity scores \citep{fong2018covariate}.
    \texttt{ps} is a normal-linear regression based propensity score model.
    \texttt{PW-boosting} is a permutation weighting model using a gradient boosted decision tree.
    \texttt{PW-glm} is a permutation weighting model using a logistic regression.
    \texttt{Unweighted} uses no weighting.
    }
    \label{fig:KS-cont-incorrect}
\end{figure}

Figure \ref{fig:KS-cont-correct} shows the results when treatment is a linear function of the observed covariates.
Propensity score weighting does quite well in terms of reducing bias (consistently with the lowest amount of bias out of all methods), but permutation weighting (particularly using a boosted model) does a better job of trading off bias and variance so as to reduce IRMSE (outperforming the normal-linear IPSW model by around 15\% at $n=2000$, for instance).
Thus, even when the propensity score model is well specified, boosting is able to outperform it by more smartly regularizing (and, thereby, reducing variance).
The level of regularization may be tuned rigorously using cross-validation.

In the case of misspecification, figure \ref{fig:KS-cont-incorrect} shows the learning curves of the various methods.
Permutation weighting outperforms all methods at all examined sample sizes in both bias and accuracy.
While PW with a logistic classifier has very similar levels of bias as does \cite{fong2018covariate}, it does so with sufficiently lower variance that even at $n=500$ it reduces IRMSE by around 15\%.
The boosting model improves upon the linear model both in terms of bias and IRMSE; at $n=2000$, boosting provides estimates with around 40\% lower IRMSE than does npCBPS.
A useful point of comparison is that permutation weighting outperforms the current state of the art by around four times as much as the state of the art improves on no weighting whatsoever.

\section{Doubly Robust Estimation}
\label{app:doublerobust}
A straightforward estimator for the dose response function is given by the so-called `direct method' (e.g. see \cite{dudik2011doubly}). For this method of estimation, the dose-response at $a$ would be estimated as:
\begin{equation*}
\hat{Y}^{DM}(a) = \int_\mathcal{X}\mu(X, a) dX
\end{equation*}

The direct method, then, trains a predictor which predicts $\mathbb{E}[Y\mid X, a]$ with $\mu(X, a)$ and then for each $a$, marginalizes over the covariate distribution $\mathcal{X}$.

It's possible to improve on the direct method by incorporating a double robust estimator as in \cite{kennedy2016doublerobust}:

\begin{equation*}
\hat{Y}_i^{DR}(a_i) = \frac{Y_i - \mu(X_i, a_i)}{\pi(a_i\mid X_i)} \cdot \int_\mathcal{X} \pi(a_i\mid X) dX + \hat{Y}^{DM}(a_i)
\end{equation*}

This quantity provides unbiased estimates when either the propensity score model or the outcome model is correctly specified. We can further swap out the propensity scores in the simple doubly-robust estimators for those generated through PW or any other IPSW-like method as in:

\begin{equation*}
\hat{Y}_i^{DR-PW}(a_i, \boldsymbol w) = (Y_i - \mu(X_i, a_i)) w_i + \hat{Y}^{DM}(a_i)
\end{equation*}

That is, we can observe that the first term of the doubly-robust estimator is simply the IPSW, and replace it with any IPSW-like weight. 
Moreover, the use of case-weights in estimating $\hat{Y}^{DM}(a)$ does nothing to dampen the consistency of the estimator (under our assumption of positivity), but with appropriate weights may do a better job of ensuring that accuracy is preferenced in areas of $\mathcal{X}$ where it is actually needed for the dose-response function.
Thus, in practice, we apply case-weights when estimating a machine learning model for use in the direct method.

In order to actually estimate a dose-response function from data, it is necessary to additionally pass these estimates for each observation through a flexible function approximation method (such as a local kernel regression method) to estimate the curve of $\mathbb{E}[Y(a)]$. We will not dwell on this latter component, and interested readers may see \cite{kennedy2016doublerobust} for more details.

\section{Extended Kang and Schafer (2007) simulation results}
\label{app:more_sims}

\begin{figure}
    \centering
\begin{center}
    \includegraphics[width=.85\textwidth]{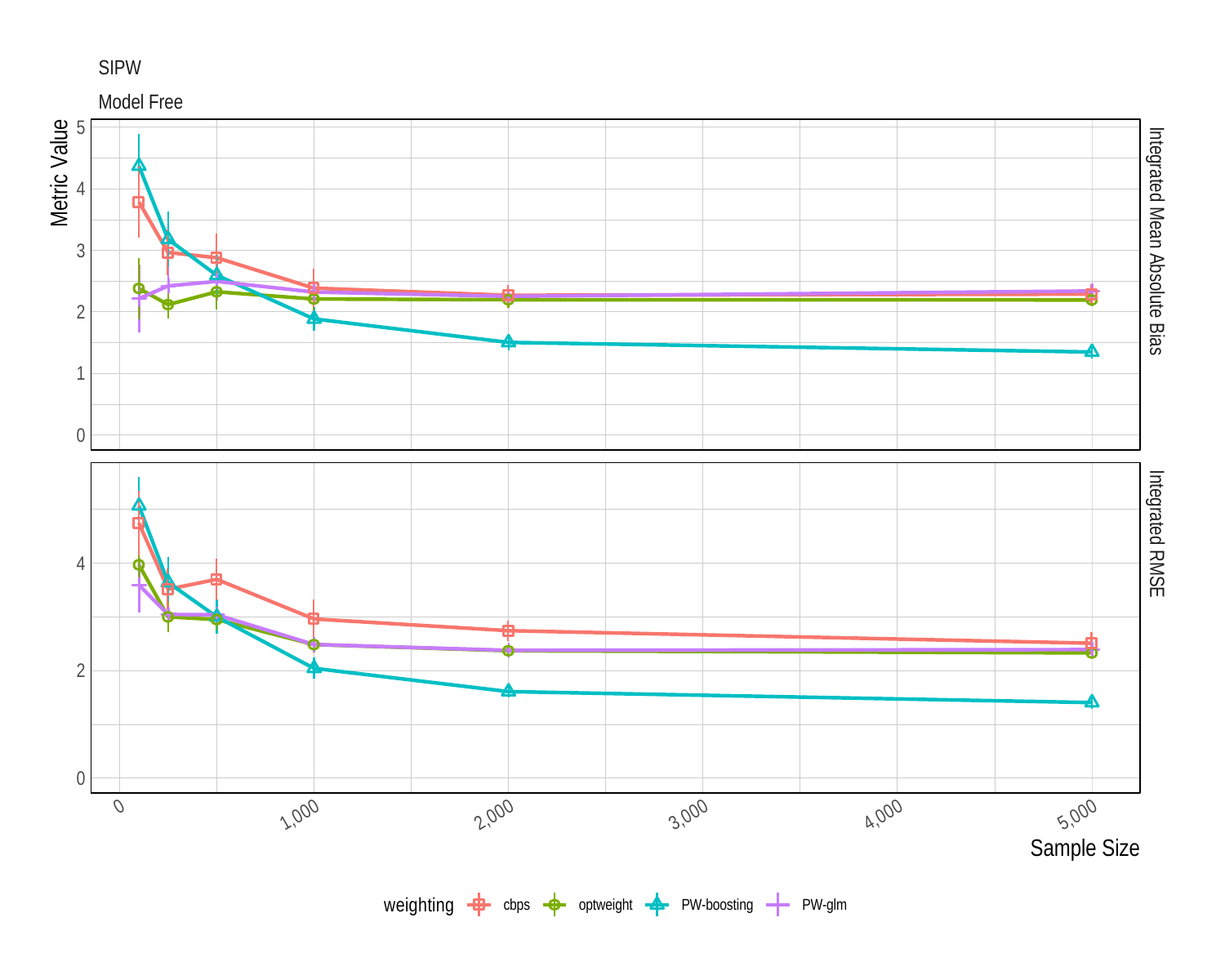}
\end{center}
\caption{}
\label{fig:KS-miss-zoom}
\end{figure}

Figure~\ref{fig:KS-miss-zoom} shows the results of the binary model when the covariates are misspecified.
In this figure, only the best performing methods are shown.

The following tables show IRMSE and Bias estimates (defined identically as in the main text) for all combinations of outcome estimation strategies (weighting only, direct method and doubly-robust), models (OLS and random forests) and weighting methods. Estimates of bias or IRMSE are followed by the standard error (estimated via non-parametric bootstrap).

\subsection{Binary treatment -- Kang and Schafer (2007)}
\footnotesize
\captionsetup[table]{labelformat=empty,skip=1pt}
\begin{longtable}{llrrrrrrr}
\caption*{
\large Well specified \\ 
\footnotesize N = 2000\\ 
} \\ 
\toprule
 & Metric & Unweighted & Logit & Boosting & CBPS & SBW & PW (Logit) & PW (Boosting) \\ 
\midrule
\multicolumn{1}{l}{IPSW} \\ 
\midrule
Model Free & Bias & 9.52 $\pm$ 0.10 & 0.45 $\pm$ 0.07 & 4.42 $\pm$ 0.06 & 0.08 $\pm$ 0.04 & 0.50 $\pm$ 0.00 & 0.67 $\pm$ 0.03 & 0.08 $\pm$ 0.05 \\ 
 & IRMSE & 9.55 $\pm$ 0.10 & 0.92 $\pm$ 0.06 & 4.46 $\pm$ 0.06 & 0.74 $\pm$ 0.03 & 0.50 $\pm$ 0.00 & 0.77 $\pm$ 0.03 & 0.52 $\pm$ 0.06 \\ 
\midrule
\multicolumn{1}{l}{Direct Method} \\ 
\midrule
OLS & Bias & 0.50 $\pm$ 0.00 & 0.50 $\pm$ 0.00 & 0.49 $\pm$ 0.00 & 0.50 $\pm$ 0.00 & 0.50 $\pm$ 0.00 & 0.50 $\pm$ 0.00 & 0.50 $\pm$ 0.00 \\ 
 & IRMSE & 0.50 $\pm$ 0.00 & 0.50 $\pm$ 0.00 & 0.49 $\pm$ 0.00 & 0.50 $\pm$ 0.00 & 0.50 $\pm$ 0.00 & 0.50 $\pm$ 0.00 & 0.50 $\pm$ 0.00 \\ 
Random Forest & Bias & 0.42 $\pm$ 0.01 & 0.03 $\pm$ 0.01 & 0.13 $\pm$ 0.01 & 0.06 $\pm$ 0.02 & 0.06 $\pm$ 0.00 & 0.03 $\pm$ 0.01 & 0.03 $\pm$ 0.00 \\ 
 & IRMSE & 0.44 $\pm$ 0.02 & 0.09 $\pm$ 0.01 & 0.15 $\pm$ 0.01 & 0.18 $\pm$ 0.02 & 0.16 $\pm$ 0.01 & 0.09 $\pm$ 0.01 & 0.08 $\pm$ 0.01 \\ 
\midrule
\multicolumn{1}{l}{Doubly Robust} \\ 
\midrule
OLS & Bias & 0.50 $\pm$ 0.00 & 0.50 $\pm$ 0.00 & 0.49 $\pm$ 0.00 & 0.50 $\pm$ 0.00 & 0.50 $\pm$ 0.00 & 0.50 $\pm$ 0.00 & 0.50 $\pm$ 0.00 \\ 
 & IRMSE & 0.50 $\pm$ 0.00 & 0.50 $\pm$ 0.00 & 0.49 $\pm$ 0.00 & 0.50 $\pm$ 0.00 & 0.50 $\pm$ 0.00 & 0.50 $\pm$ 0.00 & 0.50 $\pm$ 0.00 \\ 
Random Forest & Bias & 0.47 $\pm$ 0.02 & 0.18 $\pm$ 0.01 & 0.26 $\pm$ 0.01 & 0.18 $\pm$ 0.01 & 0.03 $\pm$ 0.01 & 0.19 $\pm$ 0.01 & 0.18 $\pm$ 0.01 \\ 
 & IRMSE & 0.50 $\pm$ 0.02 & 0.22 $\pm$ 0.01 & 0.28 $\pm$ 0.02 & 0.27 $\pm$ 0.02 & 0.16 $\pm$ 0.01 & 0.22 $\pm$ 0.01 & 0.21 $\pm$ 0.01 \\ 
\bottomrule
\end{longtable}

\captionsetup[table]{labelformat=empty,skip=1pt}
\begin{longtable}{llrrrrrrr}
\caption*{
\large Misspecified\\
\small N = 2000\\
} \\
\toprule
 & Metric & Unweighted & Logit & Boosting & CBPS & SBW & PW (Logit) & PW (Boosting) \\
\midrule
\multicolumn{1}{l}{IPSW} \\
\midrule
Model Free & Bias & 9.66 $\pm$ 0.12 & 5.87 $\pm$ 0.71 & 4.72 $\pm$ 0.06 & 2.28 $\pm$ 0.08 & 2.22 $\pm$ 0.06 & 2.27 $\pm$ 0.06 & 1.52 $\pm$ 0.05 \\
 & IRMSE & 9.68 $\pm$ 0.13 & 8.04 $\pm$ 1.27 & 4.75 $\pm$ 0.07 & 2.74 $\pm$ 0.08 & 2.38 $\pm$ 0.05 & 2.38 $\pm$ 0.07 & 1.62 $\pm$ 0.06 \\
\midrule
\multicolumn{1}{l}{Direct Method} \\
\midrule
OLS & Bias & 2.78 $\pm$ 0.05 & 2.66 $\pm$ 0.08 & 1.04 $\pm$ 0.03 & 2.69 $\pm$ 0.09 & 2.22 $\pm$ 0.06 & 2.14 $\pm$ 0.05 & 1.13 $\pm$ 0.05 \\
 & IRMSE & 2.79 $\pm$ 0.06 & 2.71 $\pm$ 0.08 & 1.11 $\pm$ 0.04 & 2.75 $\pm$ 0.10 & 2.38 $\pm$ 0.05 & 2.19 $\pm$ 0.06 & 1.19 $\pm$ 0.05 \\
Random Forest & Bias & 0.78 $\pm$ 0.02 & 0.25 $\pm$ 0.01 & 0.37 $\pm$ 0.01 & 0.29 $\pm$ 0.01 & 0.23 $\pm$ 0.02 & 0.24 $\pm$ 0.01 & 0.20 $\pm$ 0.01 \\
 & IRMSE & 0.79 $\pm$ 0.02 & 0.27 $\pm$ 0.01 & 0.39 $\pm$ 0.02 & 0.33 $\pm$ 0.01 & 0.30 $\pm$ 0.01 & 0.27 $\pm$ 0.01 & 0.23 $\pm$ 0.01 \\
\midrule
\multicolumn{1}{l}{Doubly Robust} \\
\midrule
OLS & Bias & 2.78 $\pm$ 0.05 & 2.66 $\pm$ 0.08 & 1.04 $\pm$ 0.03 & 2.69 $\pm$ 0.09 & 2.22 $\pm$ 0.06 & 2.14 $\pm$ 0.05 & 1.13 $\pm$ 0.05 \\
 & IRMSE & 2.79 $\pm$ 0.06 & 2.71 $\pm$ 0.08 & 1.11 $\pm$ 0.04 & 2.75 $\pm$ 0.10 & 2.38 $\pm$ 0.05 & 2.19 $\pm$ 0.06 & 1.19 $\pm$ 0.05 \\
Random Forest & Bias & 0.90 $\pm$ 0.02 & 0.53 $\pm$ 0.02 & 0.60 $\pm$ 0.02 & 0.54 $\pm$ 0.02 & 0.39 $\pm$ 0.02 & 0.49 $\pm$ 0.02 & 0.48 $\pm$ 0.02 \\
 & IRMSE & 0.92 $\pm$ 0.02 & 0.54 $\pm$ 0.02 & 0.61 $\pm$ 0.02 & 0.56 $\pm$ 0.02 & 0.44 $\pm$ 0.02 & 0.51 $\pm$ 0.02 & 0.49 $\pm$ 0.02 \\
\bottomrule
\end{longtable}

\newpage
\subsection{Continuous treatment -- Kang and Schafer (2007)}

\captionsetup[table]{labelformat=empty,skip=1pt}
\begin{longtable}{llrrrrr}
\caption{
\large  Well specified\\ 
\small N = 2000\\ 
} \\ 
\toprule
model & Metric & Unweighted & Normal-Linear & npCBPS & PW (Logit) & PW (Boosting) \\ 
\midrule
\multicolumn{1}{l}{IPSW} \\ 
\midrule
Model Free & Bias & 15.750 $\pm$ 0.114 & 4.113 $\pm$ 0.285 & 7.996 $\pm$ 0.377 & 8.862 $\pm$ 0.116 & 6.021 $\pm$ 0.149 \\ 
 & IRMSE & 16.142 $\pm$ 0.114 & 8.685 $\pm$ 0.242 & 11.445 $\pm$ 0.191 & 9.697 $\pm$ 0.119 & 7.576 $\pm$ 0.143 \\ 
\midrule
\multicolumn{1}{l}{Direct Method} \\ 
\midrule
OLS & Bias & 0.269 $\pm$ 0.000 & 0.269 $\pm$ 0.000 & 0.269 $\pm$ 0.000 & 0.269 $\pm$ 0.000 & 0.269 $\pm$ 0.000 \\ 
 & IRMSE & 0.269 $\pm$ 0.000 & 0.284 $\pm$ 0.003 & 0.298 $\pm$ 0.009 & 0.270 $\pm$ 0.000 & 0.273 $\pm$ 0.001 \\ 
Random Forest & Bias & 2.507 $\pm$ 0.031 & 1.171 $\pm$ 0.019 & 1.444 $\pm$ 0.026 & 1.302 $\pm$ 0.014 & 1.280 $\pm$ 0.016 \\ 
 & IRMSE & 2.574 $\pm$ 0.030 & 1.229 $\pm$ 0.021 & 1.499 $\pm$ 0.028 & 1.339 $\pm$ 0.015 & 1.321 $\pm$ 0.017 \\ 
\midrule
\multicolumn{1}{l}{Doubly Robust} \\ 
\midrule
OLS & Bias & 0.249 $\pm$ 0.001 & 0.255 $\pm$ 0.003 & 0.259 $\pm$ 0.002 & 0.260 $\pm$ 0.001 & 0.257 $\pm$ 0.002 \\ 
 & IRMSE & 0.272 $\pm$ 0.002 & 0.354 $\pm$ 0.008 & 0.372 $\pm$ 0.017 & 0.289 $\pm$ 0.001 & 0.304 $\pm$ 0.003 \\ 
Random Forest & Bias & 2.620 $\pm$ 0.033 & 1.423 $\pm$ 0.021 & 1.719 $\pm$ 0.027 & 1.603 $\pm$ 0.015 & 1.560 $\pm$ 0.017 \\ 
 & IRMSE & 2.711 $\pm$ 0.031 & 1.500 $\pm$ 0.025 & 1.797 $\pm$ 0.029 & 1.663 $\pm$ 0.017 & 1.624 $\pm$ 0.019 \\ 
\bottomrule
\end{longtable}

\captionsetup[table]{labelformat=empty,skip=1pt}
\begin{longtable}{llrrrrr}
\caption{
\large Misspecified\\ 
\small N = 2000\\ 
} \\ 
\toprule
model & Metric & Unweighted & Normal-Linear & npCBPS & PW (Logit) & PW (Boosting) \\ 
\midrule
\multicolumn{1}{l}{IPSW} \\ 
\midrule
Model Free & Bias & 15.549 $\pm$ 0.130 & 16.810 $\pm$ 0.526 & 10.821 $\pm$ 0.259 & 10.810 $\pm$ 0.126 & 8.406 $\pm$ 0.141 \\ 
 & IRMSE & 16.002 $\pm$ 0.115 & 23.581 $\pm$ 0.881 & 14.747 $\pm$ 0.315 & 11.637 $\pm$ 0.143 & 9.418 $\pm$ 0.150 \\ 
\midrule
\multicolumn{1}{l}{Direct Method} \\ 
\midrule
OLS & Bias & 0.269 $\pm$ 0.000 & 1.351 $\pm$ 0.676 & 0.551 $\pm$ 0.195 & 2.539 $\pm$ 0.043 & 1.276 $\pm$ 0.053 \\ 
 & IRMSE & 0.269 $\pm$ 0.000 & 4.436 $\pm$ 2.533 & 1.687 $\pm$ 0.178 & 2.549 $\pm$ 0.049 & 1.323 $\pm$ 0.051 \\ 
Random Forest & Bias & 3.221 $\pm$ 0.034 & 6.865 $\pm$ 3.195 & 2.601 $\pm$ 0.051 & 2.544 $\pm$ 0.036 & 2.725 $\pm$ 0.056 \\ 
 & IRMSE & 3.273 $\pm$ 0.032 & 24.459 $\pm$ 13.162 & 2.685 $\pm$ 0.056 & 2.600 $\pm$ 0.038 & 2.811 $\pm$ 0.059 \\ 
\midrule
\multicolumn{1}{l}{Doubly Robust} \\ 
\midrule
OLS & Bias & 2.450 $\pm$ 0.050 & 1.709 $\pm$ 0.461 & 1.326 $\pm$ 0.105 & 2.382 $\pm$ 0.044 & 1.306 $\pm$ 0.059 \\ 
 & IRMSE & 2.879 $\pm$ 0.061 & 6.012 $\pm$ 1.929 & 3.922 $\pm$ 0.211 & 2.692 $\pm$ 0.045 & 2.285 $\pm$ 0.047 \\ 
Random Forest & Bias & 3.427 $\pm$ 0.034 & 6.922 $\pm$ 3.121 & 2.896 $\pm$ 0.051 & 2.840 $\pm$ 0.037 & 3.027 $\pm$ 0.056 \\ 
 & IRMSE & 3.503 $\pm$ 0.033 & 24.528 $\pm$ 13.104 & 2.994 $\pm$ 0.056 & 2.918 $\pm$ 0.039 & 3.130 $\pm$ 0.059 \\ 
\bottomrule
\end{longtable}

\section{Extended LaLonde simulation results}
\label{app:lalonde_extra}

\begin{figure}
    \centering
    \includegraphics[width=\textwidth]{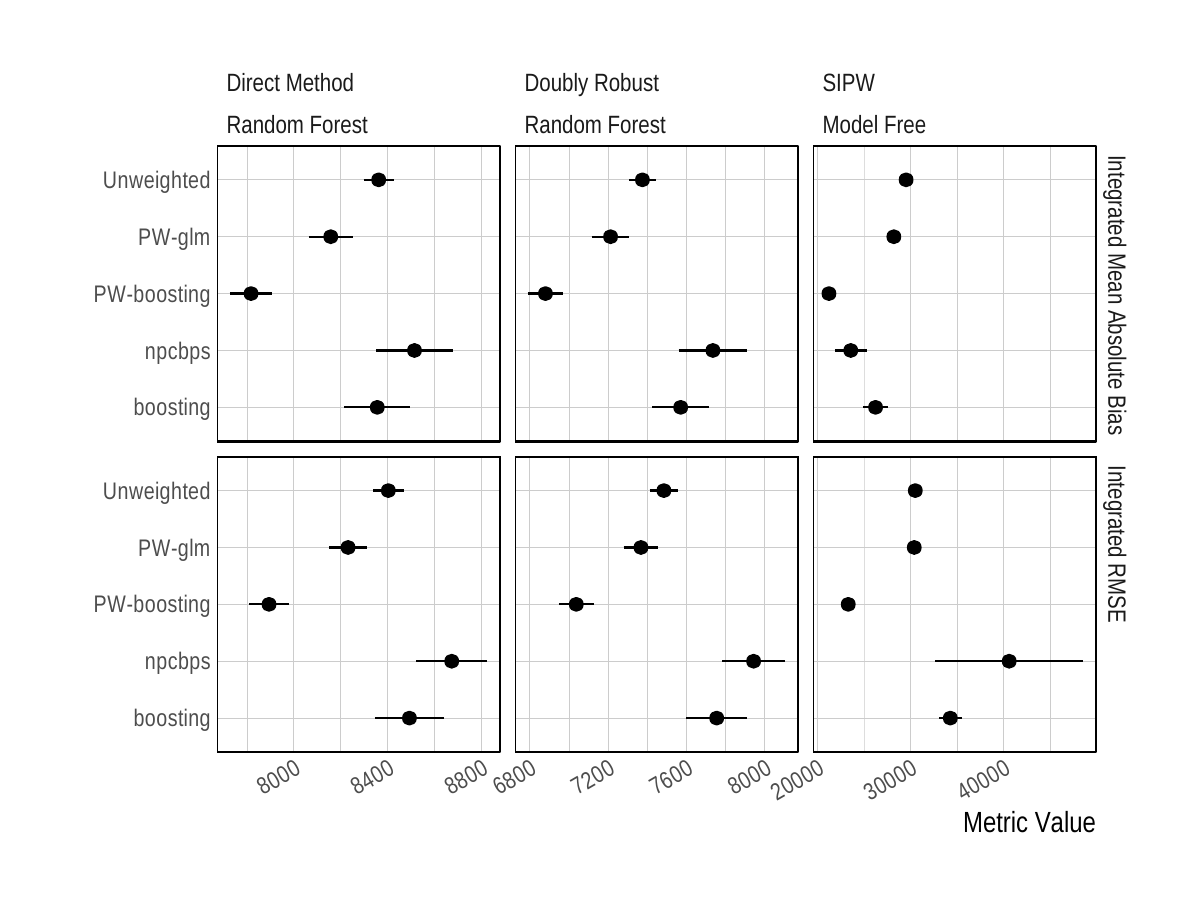}
    \caption{This figure shows additional results for the LaLonde simulation which incorporate Random Forest outcome models.}
    \label{fig:lalonde_extra}
\end{figure}

Figure~\ref{fig:lalonde_extra} provides additional results for the LaLonde data generating process.
In these results, we show bias and accuracy for a weighting-only model (replicated from the main body), a direct-method model (incorporating case-weights generated by the various methods) and a \citep{kennedy2016doublerobust} style double robust estimator. 
In these plots, a linear propensity score model is left off of the plots as it performs so much more poorly that the x-axis is greatly skewed.

\end{document}